\documentclass[structabstract]{aa}  
\usepackage{graphicx}
\usepackage{longtable}
\usepackage{enumerate}
\usepackage[authoryear]{natbib}
\bibpunct{(}{)}{;}{a}{}{,}
\usepackage{txfonts}
\begin{document}
\title{Red supergiants around the obscured open cluster Stephenson~2\fnmsep\thanks{Based on observations 
collected at the William Herschel Telescope (La Palma, Spain)}}
\author{
I.~Negueruela\inst{1}
\and A.~Marco\inst{1}
\and C.~Gonz\'alez-Fern\'andez\inst{1}
\and F.~Jim\'enez-Esteban\inst{2,3,4}
\and J.~S.~Clark\inst{5}
\and M.~Garcia\inst{6,7}
\and E.~Solano\inst{2,3}
}

\institute{
Departamento de F\'{\i}sica, Ingenier\'{\i}a de Sistemas y
  Teor\'{\i}a de la Se\~{n}al, Universidad de Alicante, Apdo. 99, E-03080
  Alicante, Spain\\
\email{ignacio.negueruela@ua.es}
\and
Centro de Astrobiolog\'{\i}a (INTA-CSIC), Departamento de Astrof\'{\i}sica,  PO Box 78, E-28691, Villanueva de la Ca\~nada, Madrid, Spain
\and
Spanish Virtual Observatory, Spain 
\and
Saint Louis University, Madrid Campus, Division of Science and Engineering, Avda.~del Valle 34, E-28003 Madrid, Spain
\and
Department of Physics and Astronomy, The Open 
University, Walton Hall, Milton Keynes, MK7 6AA, UK
\and 
Instituto de Astrof\'{\i}sica de Canarias, E-38200 La Laguna, Tenerife, Spain
\and 
Departamento de Astrof\'{\i}sica, Universidad de La Laguna (ULL), E-38206 La Laguna, Tenerife, Spain
}

\abstract{Several clusters of red supergiants have been discovered in a small region of the Milky Way close to the base of the Scutum-Crux Arm and the tip of the Long Bar. Population synthesis models indicate that they must be very massive to harbour so many supergiants. Amongst these clusters, Stephenson~2, with a core grouping of 26 red supergiants, is a strong candidate to be the most massive cluster in the Galaxy.}
{Stephenson~2 is located close to a region where a strong over-density of red supergiants had been found. We explore the actual cluster size and its possible connection to this over-density.}
{Taking advantage of Virtual Observatory tools, we have performed a
  cross-match between the DENIS, USNO-B1 and 2MASS catalogues to identify candidate obscured luminous red stars around Stephenson~2, and in a control nearby region. More than 600 infrared bright stars fulfill the colour criteria, with the vast majority having a counterpart in the $I$ band and $>400$ being sufficiently bright in $I$ to allow observation with a 4-m class telescope. We have observed a subsample of $\sim$250 stars, using the multi-object, wide-field, fibre spectrograph AF2 on the WHT telescope in La Palma, obtaining intermediate-resolution spectroscopy in the 7500--9000\AA\ range. We derive spectral types and luminosity classes for all these objects and measure their radial velocities.}   
{Our targets turn out to be G and K supergiants, late ($\geq$M4) M giants, and M-type bright giants (luminosity class II) and supergiants. We find $\sim$35 red supergiants with radial velocities similar to Stephenson~2 members, spread over the two areas surveyed. In addition, we find $\sim$40 red supergiants with radial velocities incompatible in principle with a physical association.}
{Our results show that Stephenson~2 is not an isolated cluster, but part of a huge structure likely containing hundreds of red supergiants, with radial velocities compatible with the terminal velocity at this Galactic longitude (and a distance $\sim$6~kpc). In addition, we find evidence of several populations of massive stars at different distances along this line of sight.} 

\maketitle

\section{Introduction}

In the past few years, a flurry of discoveries has revealed several clusters of red supergiants (RSGs) located in a small region of the Galactic plane,
 between $l=24\degr$ and $l=29\degr$ \citep[e.g.,][]{figer06, davies07,
   clark09,neg11}. These clusters are so heavily obscured  that, until now, the only members observed are the RSGs. Population synthesis models suggest that the clusters must contain very large stellar populations to harbour so many RSGs  \citep[e.g.,][]{davies07}. Recent simulations indicate that, at typical ages $\tau=10-20\:$Myr \citep{davies07,davies08}, one should expect $\sim$$10^{4}\,M_{\sun}$ in stars for each 3--5 RSGs \citep{simonw51}. 

\defcitealias{davies07}{D07}
\defcitealias{neg11}{Paper~I}

 RSGC2 = Stephenson~2 (Ste~2; $l=26\fdg2$, $b=-0\fdg1$) is the least obscured and apparently most massive of the red supergiant clusters. Discovered by \citet{stephenson}, the cluster was found to have a clump of RSGs \citep{ortolani}. \citet[henceforth D07]{davies07} obtained $K$-band spectroscopy of a large number of bright sources within $7\arcmin$ of the nominal cluster centre, finding more than 25 RSGs that shared similar radial velocities. After defining an average radial velocity for the cluster and eliminating outliers, they found $v_{{\rm LSR}}=+109.3\pm0.7\:{\rm km}\,{\rm s}^{-1}$, with the uncertainty representing Poisson statistics for 26 likely members with measured values \citepalias{davies07}. Stars were considered members if their $v_{{\rm LSR}}$ was within $\pm10\:{\rm km}\,{\rm s}^{-1}$ of the average value. On the other hand, \citet{deguchi} obtain $v_{{\rm LSR}}$\,$=+96.5\:{\rm km}\,{\rm s}^{-1}$ from measurements of SiO masers associated with four cluster members. They attribute the difference to systematic effects.

The age of Ste~2 has been estimated at  $\tau$\,$=17\pm3$~Myr \citepalias{davies07}, consistent with a value $\sim$20~Myr derived by \citet{ortolani}. Assuming that the 26 RSGs are members, \citetalias{davies07} estimate, using population synthesis models, that the underlying cluster population must be $M_{{\rm cl}}\ga5\times10^{4}\,M_{\sun}$. Conversely, \citet{deguchi} claim that the spatial distribution of RSGs argues for the existence of two clusters, which they call Ste~2 and Ste~2 SW, though their radial velocities do not seem to be significantly different. In summary, the actual spatial extent, and hence total mass, of Ste~2 is still poorly determined.

The line of sight to the RSG clusters (RSGCs) passes through several Galactic arms and reaches an area of very high obscuration at a distance comparable to those of the clusters \citep{neg10}. Stellar densities are extremely high at moderate $K$ magnitudes, and membership of individual stars is difficult to assess. \citetalias{davies07} found a large population of field stars with $K_{{\rm S}}$ magnitudes comparable to the supergiants. Indeed, down to their limiting magnitude $K_{{\rm S}}=8$, there are many more field stars than RSG members in the $r=7\arcmin$ circle analysed (though RSGs are more numerous than field stars for $K_{{\rm S}}<6.2$).  \citetalias{davies07} identify these field stars as predominantly foreground giants, though several putative RSGs with $v_{{\rm rad}}$ suggestive of non-membership were found.

 In a recent paper (\citealt{neg11}; from now on, \citetalias{neg11}), we studied the spatial extent of another cluster of red supergiants, RSGC3, with estimated $\tau$=16--20$\,$Myr and an inferred $M_{{\rm cl}}$=2--4$\times10^4\,M_{\sun}$  \citep{clark09,alexander09}, finding that it is part of an association, which includes smaller clusters of red supergiants. This association contains $\ga 50$ RSGs and hence a total mass $\ga 10^{5}\,M_{\sun}$. This size has been confirmed (and likely surpassed) by the recent discovery of another cluster of RSGs in its vicinity \citep{carlos}.

The connection of the RSGC3 association to Ste~2 and the other RSG clusters is unclear. Though all the clusters are located at a similar distance and have similar ages, they span $\sim$500~pc in projection. There have been suggestions that this
Scutum Complex represents a giant star formation region triggered by the dynamical excitation of the Galactic Bar, whose tip is believed to intersect the Scutum-Crux Arm close to this region
(see discussion in \citetalias{davies07}; \citealt{garzon}). The possibility of a hierarchical star-forming structure extending over such a long distance seems unlikely as it would be larger than the largest cluster complexes seen in external galaxies \citep[for instance, $\sim240$~pc in M51;][]{bastian}, even though less coherent structures \citep[aggregates in the terminology of][]{efremov04} are seen spanning comparable distances.

 All the RSG clusters have similar ages, in the 10--20~Myr range, but this does not guarantee an association. RSGC1, with an age $\sim$12~Myr \citep{davies08}, could be more closely allied to its close neighbour, the Quartet cluster, with an age between 3--8~Myr \citep{messi09}, while the RSGC3 association lies very close in the sky to the W43 star-forming complex, also at $d$\,$\sim6$~kpc, which covers the $l=29\fdg5-31\fdg5$ range \citep{nguyen11}.

The distribution of Galactic \ion{H}{ii} regions shows a marked maximum towards $l\sim30\degr$--$31\degr$ \citep{bania, anderson11}, where tens of sources display radial velocities in the $v_{{\rm LSR}}\approx90-105\:{\rm km}\,{\rm s}^{-1}$ interval. This is generally interpreted as the tangent point of the Scutum-Crux arm. There is a secondary maximum towards $l\sim24\degr$, where many sources have $v_{{\rm LSR}}\approx110\:{\rm km}\,{\rm s}^{-1}$, similar to the systemic velocity of RSGC1 \citep[$l$\,$=25\fdg3$; $v_{{\rm LSR}}=123\:{\rm km}\,{\rm s}^{-1}$;][]{davies08}. There is a local minimum towards $l\sim26\degr$, though the number of \ion{H}{ii} regions in that direction is still high \citep{anderson11}. Galactic molecular clouds seem to follow a similar distribution \citep[e.g.,][]{rathborne09}.

In this paper, we study the distribution of luminous red stars in the vicinity of Ste~2. Though the cluster core is clearly defined by a strong concentration of bright infrared sources, its actual extent is poorly determined. The spatial distribution of known members is very elongated, and the possibility that the cluster is surrounded by a halo has not been studied. Likewise, the fact that RSGC3 is surrounded by an extended association opens the possibility that Ste~2 is not isolated either.

Following the strategy outlined in \citetalias{neg11}, we have used a multi-object spectrograph with high multiplexing capability to obtain intermediate-resolution spectra of a large sample of bright infrared sources in the spectral region around the \ion{Ca}{ii} triplet. In Section~\ref{sec:target}, we discuss the photometric criteria used for selection of candidates and describe the observations of a sample of these candidates in the region around Ste~2 and a second sample in a control area about $1\degr$ away. In Section~\ref{sec:calib}, we summarise the analysis methodology. We have then used the criteria and calibrations developed in \citetalias{neg11} to assign spectral types and calculate colour excesses. We also measure radial velocities for all our sources. The results of the analysis are presented in Section~\ref{sec:res}. Finally, we discuss their implications in Section~\ref{sec:discu} and wrap up with the conclusions.

\section{Target selection and observations}
\label{sec:target}

 In \citetalias{neg11} we showed that a careful application of photometric criteria over existing catalogues could select obscured red luminous stars with high accuracy. Unfortunately, the photometric properties of RSGs do not allow a clear separation from M-type giants, implying the need for spectroscopic classifications. 

\begin{figure}
\resizebox{\columnwidth}{!}{
\includegraphics[angle=-90,clip]{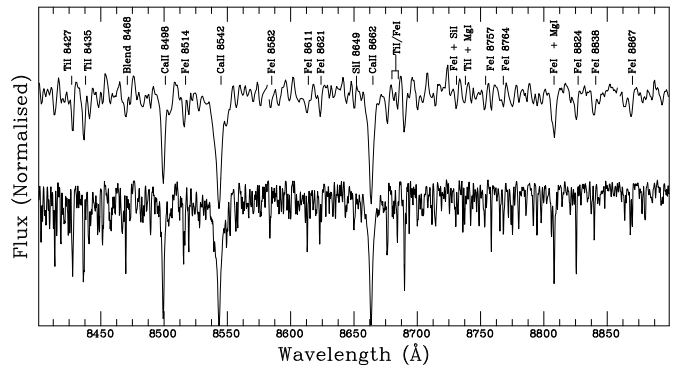}}
\caption{The spectrum of one of our targets (J18414143$-$0531378) observed with AF2 (top), at $R=4\,100$ and with ISIS at $R\approx16\,000$. At the AF2 resolution, many of the lines seen in the ISIS spectrum are blended into broad features. An interesting example is the 8514\AA\ blend (mainly \ion{Fe}{i}) that appears as a single feature at the AF2 resolution. This feature is very strong in luminous supergiants (see Fig.~\ref{fig:lums}). \label{comparison}}
\end{figure}

\subsection{Selection of candidates}

Three catalogues have been used to search for candidates. We have
started with the 2MASS Point Source Catalogue \citep{skru06}. 
2MASS contains photometric data for 471 million sources in the $J$ (1.25\,$\mu$m), $H$ (1.65\,$\mu$m), and $K_{{\rm S}}$
(2.16\,$\mu$m) near-infrared bands. Typical photometric and astrometric
uncertainties of 2MASS are less than 0.03~mag and 100~mas,
respectively.  Three selection conditions were applied on these
data, based on the criteria discussed in \citetalias{neg11}: 
\begin{enumerate}
\item We selected stars bright in the $K_{{\rm S}}$ band: $K_{{\rm S}}\leq7.0$.
\item We selected reddened stars by requiring a colour redder than that of any star: $(J-K_{{\rm S}})>1.3$.
\item We selected stars with infrared reddening-free parameter, $Q_{{\rm IR}}=(J-H)-1.8\times(H-K_{{\rm S}})$, typical of RSGs:
$0.1\leq Q_{{\rm IR}}\leq0.4$. Though some RSGs have $Q_{{\rm IR}}$  outside this range \citep[cf.][]{messineo12}, this cut selects most RSGs, while leaving out low-luminosity red stars.
\end{enumerate}  
The vast majority of our targets present a good quality flag (Qflg = A, B, C, or D) in all three 2MASS bands. A few objects with other quality flags were included in the sample and assigned low priority, after we checked out that they were not badly blended. In the end, only two such objects were observed, J18402969$-$0546359\fnmsep\footnote{Throughout this paper, 2MASS identifications are given with coordinates only, to save space.} and J18403226$-$0611584.

The 2MASS objects selected were subsequently cross-matched with two others catalogues, DENIS\fnmsep\footnote{{\tt http://cdsweb.u-strasbg.fr/denis.html}}
and USNO-B.1 \citep{monet03}. DENIS provides photometry in the Gunn $i$ band (0.8\,$\mu$m), while USNO-B.1 gives magnitudes in a photographic near-infrared ``$I$'' band. We applied two magnitude criteria to the objects cross-matched: i) we rejected all stars bright in $I$ ($I<10$), so as to leave out unreddened stars; and ii) we required $I\leq17$, to select only stars observable with current instrumentation. We found more than 600 candidates within $1\degr$ of the centre of Ste~2 in the 2MASS/DENIS cross-match, and a slightly smaller number in the 2MASS/USNO cross-match. Around 30\% of the USNO sources do not have counterparts in DENIS, resulting in slightly above 800 unique candidates. These are not all the candidate red luminous stars in the area, but only those with a counterpart in the far red with $10\leq I \leq 17$. 

For the AF2 observations presented in this paper, we had to restrict the target list to objects brighter than $I=15.1$, to make sure that the star would give a usable spectrum with a 4~m class telescope.  Using this new criterion, the main target area (defined below and covering $\sim0.8$~deg$^2$) contains 226 2MASS/DENIS cross-matches, and 202 2MASS/USNO cross-matches, with 246 unique objects.


To apply the colour-magnitude criteria and to cross-match the catalogues, we took advantage of the tools offered by the Virtual Observatory (VO), an international project designed to provide the astronomical community with the data access and the research tools necessary to enable the scientific exploration of the digital, multi-wavelength universe resident in the astronomical data archives. In particular, we used Aladin \citep{bonnarel}, a VO-compliant interactive sky atlas that allows users to visualise and analyse digitised astronomical images, astronomical catalogues, and databases available from the VO services. We also used TOPCAT\fnmsep\footnote{\tt http://www.star.bris.ac.uk/$\sim$mbt/topcat/}, another VO-tool, to purge duplicated data. TOPCAT is an interactive graphical viewer and editor for tabular data that allows the user to examine, analyse, combine and edit astronomical tables.

\subsection{Observations}

Observations were carried out with the AutoFib2+WYFFOS (AF2) multi-object, wide-field, fibre spectrograph mounted on the Prime Focus of the 4.2~m William Herschel Telescope (WHT), in La Palma, Spain. The observations were taken on the nights of 2009, June 5th (in service mode) and June 6th-7th (in visitor mode). AF2 uses 150 science fibres with diameter $1\farcs6$, and 10 fiducial bundles for acquisition and guiding. Fibres can be positioned over a field of radius $30\arcmin$, though there is  some vignetting starting at $r=20\arcmin$, and image quality outside $r=25\arcmin$ is quite poor. 
Due to the size of the fibre buttons in AF2, fibres cannot be placed closer than approximately $25\arcsec$ from each other. Because of this, it is not an appropriate instrument to observe dense clusters. 

We used the R1200R grating, blazed at 7200\AA. This grating provides a nominal dispersion of 0.32\AA/pixel over 1300\AA\ (this is the range common to spectra from all fibres; spectra from different fibres can be shifted up to about 5\% of the range, depending on their position in the field). When used in combination with the $1\farcs6$ fibres, it gives a resolution element of 2\AA. We used central wavelength 8300\AA. Our common spectral range is thus 7650-8950\AA, at resolving power $R\ga4\,000$.

\begin{figure}
\resizebox{\columnwidth}{!}{
\includegraphics[angle=0,clip]{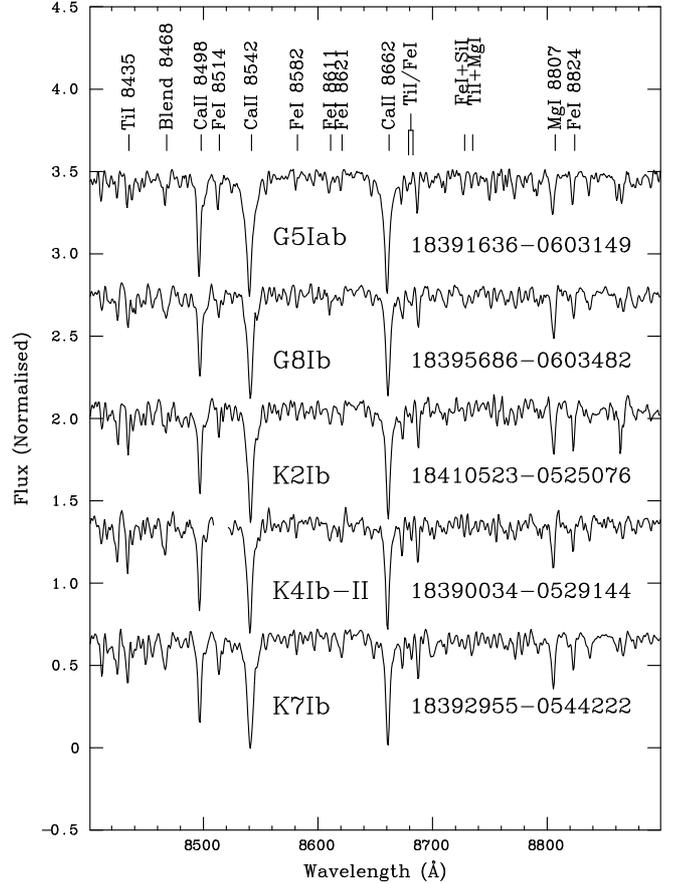}}
\caption{A sample of G--K low-luminosity supergiants observed with AF2. Note the evolution of the \ion{Ti}{i}~8683\AA/\ion{Fe}{i}~8679\AA\ ratio with spectral type. All these objects have relatively low extinction. In most of them (but not in J18392955$-$0544222) the 8620\AA\ DIB, blended with \ion{Fe}{i}~8621\AA, is hardly noticeable.\label{fig:early}}
\end{figure}

\begin{figure}[ht!]
\resizebox{\columnwidth}{!}{
\includegraphics[angle=0,clip]{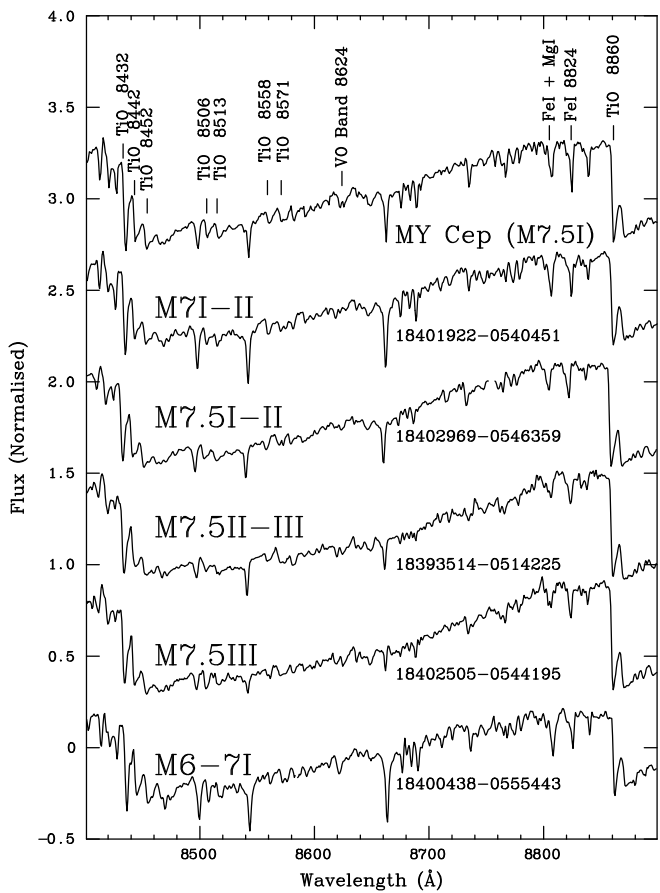}}
\caption{A sample of late M-type stars, illustrating the characteristics used for classification. The depth of the TiO bandhead at 8860\AA\ is expected to depend linearly with $T_{{\rm eff}}$. The 8620\AA\ DIB is blended with the VO bandhead at 8624\AA, only visible in stars later than M6. The spectrum of the M7.5\,I RSG MY~Cep, in the open cluster NGC~7419, is shown for comparison. All the spectra have been normalised in the same, rather arbitrary, way. \label{lates}
}
\end{figure}

AF2 uses as detector a chip mosaic, consisting of two EEV-42-80 thinned and AR coated CCDs butted along their long axis to provide a $4096\times4096$~pixel detector. As the cameras are thinned, they suffer from heavy fringing in the far red spectral region ($\sim$40\%). This effect, however, can be corrected with the observation of internal flat-fields. As AF2 is a bench-mounted spectrograph, the fringing pattern is very stable. We obtained flats at each pointing position to improve fringing removal. Each observation was divided into 5--6 exposures. Flat fields were taken immediately before and after each of the exposures. With this strategy, we have been able to reduce the fringing level to $\la$2\% for all targets with reasonably bright spectra ($\ga 4000$ counts/pixel). Except in a few spectra with low count levels, the fringing pattern is not visible in the spectra.

The sky spectra in the far-red are dominated by strong airglow emission lines. For a better sky removal, we used the ``beam-switch'' strategy (moving the telescope back and forth between objects and neighbouring sky). For each exposure, we moved the telescope $\sim$$3\arcmin$ in a random direction and took a sky exposure with the same integration time as each of the individual target exposures. This procedure was used during the nights observed in visitor mode, but not for the service observation. Though it is time-consuming, it has allowed a very good sky subtraction in most cases. Again, the worst results are found for the faintest targets. The observations from June 5th have had the sky subtraction carried out with spectra obtained through a high number of fibres allocated to blank sky positions. Though the procedure is in most cases satisfactory, it gives poorer results than the  ``beam-switch'' strategy for the faintest sources. This is due to the very different throughputs of the fibres. A few of the spectra (with $\la 3000$ counts/pixel -- meaning $\la 500$ counts/pixel in an individual exposure) were unusable due to a combination of sky and fringing contamination.

To observe Stephenson~2, the field of AF2 was centred on RA: 18h 40m 25.0s, Dec: $-05\degr 44\arcmin 30\arcsec$. This is not the nominal centre of the cluster, but lies $\sim$$15\arcmin$ to the Northeast, leaving the cluster in one of the AF2 quadrants, and extending the field in the direction towards the area where \citet{garzon} had observed an over-density of RSGs. Three configurations were observed centred on these coordinates. The first configuration contained only bright targets, with $I<11.1$. The second configuration contained targets of intermediate brightness, with $10.9\leq I \leq13.1$. The third configuration contained faint targets with $12.9\leq I <15.1$. A small overlap in magnitudes was allowed, so that a few stars could be picked more than once by the configuration software, even though no attempt was made to ensure this overlap. 

As a further check, a single configuration with intermediate exposure times was observed, displaced a few arcmin further Northeast from the main position. The area covered by this configuration and the three exposures of the Stephenson~2 field represent the ``main target area'' (see Fig.~\ref{fig:distsg}). Finally, to assess the singularity of the region surrounding Ste~2, we observed a control field, located $\sim$$1\fdg1$ away from the centre of our main field, and so not overlapping with it. The control field was chosen to be in an area where no over-density of RSGs had been claimed before. In addition, it contains a star cloud, an area with a very high density of objects, likely caused by low foreground extinction. The densest part of this star cloud is identifiable as the cluster candidate Andrews-Lindsay~5 \citep{al67}. In this control region, the selection criteria were loosened in order to probe a larger region of the parameter space. We observed a few stars with $7.0\leq K_{{\rm S}}\leq7.5$, and we allowed a few objects as faint as $I=16.1$ to be included. 

Due to the small overlap in magnitude between the three exposures, and the observation of an overlapping field, $\sim$20\% of the targets have more than one spectrum (with 10 objects having three spectra). This has allowed a good estimation of the precision of spectral type classifications, of the impact of signal-to-noise ratio on the reduction procedure and of the radial velocity determination (see Sect.~\ref{radvels}). In total, we have 233 unique stars with usable spectra. Of them, 73 lie in the control region, with the rest belonging to the main target area.

In addition, we observed a few sources using the red arm of the ISIS double-beam spectrograph, also mounted on the WHT. The observations were carried out on 2010 June 1st using the R1200R grating centred on 8700\AA. The instrument was fitted with the {\it Red+} CCD, which has very low fringing in this range, and allows reduction of the fringing to $<1\%$ simply with the use of internal lamp flats. This configuration covered the
8380\,--\,9100\AA\ range in the unvignetted section of the CCD with a nominal dispersion of 0.26\AA/pixel. We used a $0\farcs9$ slit, resulting in a resolving power, measured on the arc lamps, $R\approx16\,000$.

The original purposes of this observation were improving the radial velocity determination, assessing the impact of resolution on spectral classification and performing a general ``sanity check'' on our results. Unfortunately, the night was far from optimal, with heavy dust cover leading to important extinction and poor seeing ($>1\farcs5$ most of the time). In consequence, we could only observe 11 of the AF2 targets, and a few very bright candidates that had not been observed with AF2. Figure~\ref{comparison} displays the spectrum of one RSG observed with both instruments so that the effects of very different spectral resolutions can be assessed.

The AF2 data were reduced following the recipe implemented in the pipeline provided by the AF2 
webpage\footnote{http://www.ing.iac.es/astronomy/instruments/af2/reduction.html} at that time. The code follows the standard steps for multi-fibre spectrum reduction, using 
IRAF\footnote{IRAF is 
distributed by the National Optical Astronomy Observatory, which is operated by the Association of Universities for 
Research in Astronomy, Inc., under cooperative agreement with  the National Science Foundation.} 
tasks for bias and flat-field correction, throughput calibration between fibres, wavelength calibration and sky subtraction. The ISIS data were reduced using standard IRAF tasks for bias and flat-field correction, wavelength calibration and sky subtraction.

\section{Spectral classification}
\label{sec:calib}

\begin{figure}
\resizebox{\columnwidth}{!}{
\includegraphics[angle=0,clip]{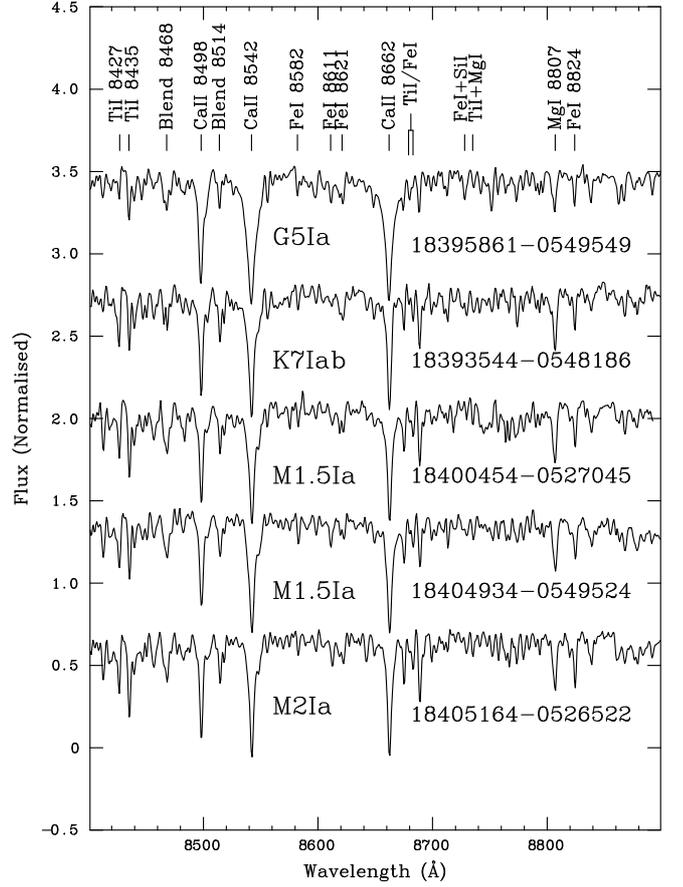}}
\caption{A sample of high-luminosity supergiants observed with AF2. All these objects suffer moderately high to high extinction, as indicated by the 8620\AA\ DIB blended with \ion{Fe}{i}~8621\AA. All these objects have radial velocities similar to members of Ste~2, except 18395861$-$0549549, which has a slightly lower $v_{{\rm rad}}$. The strength of the 8514\AA\ feature compared to the neighbouring \ion{Ti}{i}~8518\AA\ is an indicator of high luminosity.\label{fig:lums}}
\end{figure}

The classification of red luminous stars in the far red region was discussed by \citet{kh45}, \citet{sharpless} and \citet{solf} for photographic spectrograms. More recently, \citetalias{neg11} explored several criteria  presented in works discussing digital spectra \citep[e.g.,][]{ginestet94,carquillat97,diaz89,zhou91}. Here we use the most relevant results found to classify our targets. Representative spectra are presented in Fig.~\ref{fig:early}, \ref{lates}, \ref{fig:fores}, and~\ref{fig:lums}.

\subsection{Temperature determination and spectral types}
\label{sec:teff}

\begin{figure}
\resizebox{\columnwidth}{!}{
\includegraphics[angle=0,clip]{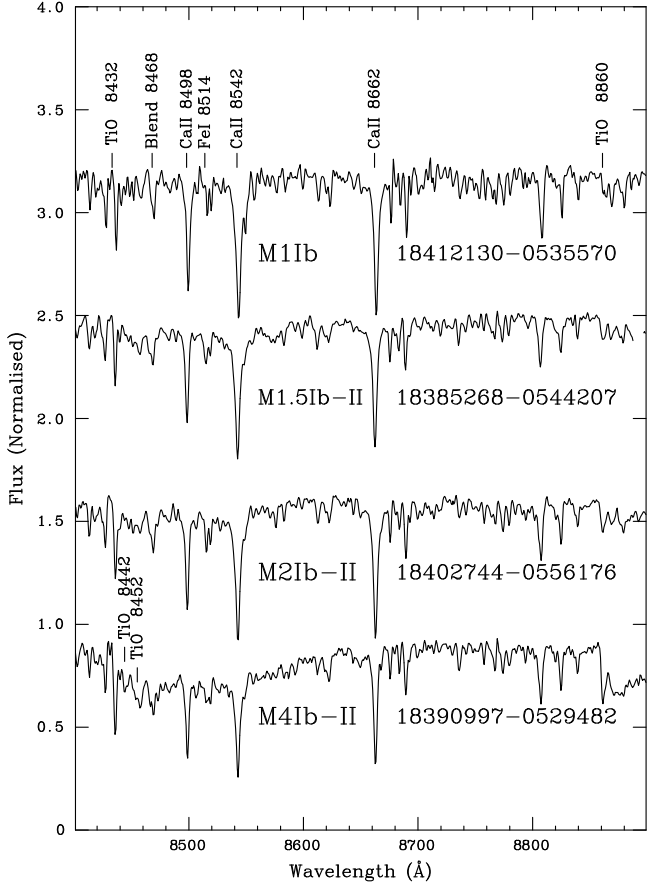}}
\caption{A sample of low-luminosity M-type supergiants observed with AF2. All these objects are likely foreground to Ste~2, as they have lower $v_{{\rm LSR}}$ and lower $E(J-K_{{\rm S}})$ colour excess. In these sources, \ion{Fe}{i}~8621\AA\ is stronger than \ion{Fe}{i}~8611\AA, indicating an important contamination by the 8620\AA\ DIB. The main luminosity indicators are the \ion{Ca}{ii} lines and the 8648\AA\ blend, but note also the weakness of the 8514\AA\ feature compared to objects in Fig.~\ref{fig:lums}. The growing strength of the TiO 8432\AA\ bandhead, blended with \ion{Ti}{i}~8435\AA, can be seen to correlate very well with TiO~8660\AA\ (the \ion{Ti}{i}~8427\AA\ line, immediately to its left, can be used as reference).
\label{fig:fores}}
\end{figure}

The far-red spectra of M-type stars are dominated by molecular TiO absorption bands, with VO bands also becoming prominent for spectral types M6 and later \citep{kirkpatrick91}. These bands are much stronger in luminous stars (giants and supergiants) than in stars close to the main sequence. In \citetalias{neg11}, we calibrated the dependence of the depth of the TiO $\delta$(0,0) $R$-branch bandhead at 8860\AA\ with $T_{{\rm eff}}$, finding a linear relation

\begin{equation}
\label{calteff}
T_{{\rm eff}}=(3848\pm38)-(1803\pm154)\times TiO \:\:\: \left[ {\mathrm K} \right] 
\, ,
\end{equation}
valid for giants and supergiants. In this expression, $TiO$ is an index defined by
\begin{equation}
TiO=1-\frac{I_{{\rm TiO}}}{I_{{\rm cont}}}
\, ,
\end{equation}

where the intensities are measured as described in \citetalias{neg11}. As the spectral sequence is a temperature sequence, the intensity of this jump is the primary spectral type indicator for M-type stars in the range observed. The intensity of the triple TiO bandhead at 8432--42--52\AA\ is an excellent secondary indicator. Some examples can be seen in Figures~\ref{lates} and~\ref{fig:fores}. The TiO bandhead at 8432\AA\ is blended with the atomic doublet \ion{Ti}{i}~8435\AA. For spectral types earlier than M0, \ion{Ti}{i}~8435\AA\ is comparable in strength to \ion{Ti}{i}~8427\AA\ (see Fig.~\ref{fig:early}). Between M1 and M3, the former feature, blended with the bandhead, grows in strength (see Fig.~\ref{fig:fores}). By M4, the bandhead features at 8442 and 8452\AA\ become visible \citep{solf}. By M6, most of the atomic features shortwards of 8600\AA\ have been obliterated by the growing strength of TiO bands, and other bandheads, such as 8558, 8571\AA\ are clearly seen. By M7, the VO bandheads start to appear and the \ion{Ca}{ii} lines become very weak in giants \citep{solf}, as illustrated in Fig.~\ref{lates}.

The TiO bandhead at 8860\AA\ becomes first noticeable, as a weak inflection in the continuum at M1. For earlier-type stars, other criteria must be used. The most effective criterion is the ratio between the \ion{Ti}{i}~8683\AA\ and the Fe\,{\sc i}~8679\AA\ lines, which is very sensitive to temperature in the G5--K5 range \citep{carquillat97}. These two lines are not fully resolved at the resolution of the AF2 spectra, but their relative intensities can be discerned and used to provide spectral types. Using all the supergiants in \citet{cenarro01}, \citet{carquillat97} and \citet{munarit99}, we find that the ratio is not strongly affected by resolution. The \ion{Ti}{i} line is very weak in early-G stars, and grows after G3 until the two lines have similar depths at G8. The \ion{Fe}{i} line becomes very weak compared to the \ion{Ti}{i} one at K5 (see Fig.~\ref{fig:early}).
This leaves a small range K5--M0.5, where differences in the spectral range considered are very subtle. Outside this narrow range, spectra can be classified with an accuracy of $\pm$1 subtypes.

\subsection{Luminosity calibration}
\label{sec:lum}

The luminosity of late-type stars can be calibrated with several indicators. In general, the intensity of the whole metallic spectrum is dependent on luminosity. The best calibrated indicators are the equivalent widths (EW) of the two strongest lines in the \ion{Ca}{ii} triplet and the \ion{Ti}{i} dominated blend at~8468\AA\ (\citealt{diaz89}, \citealt{ginestet94}, \citetalias{neg11}), which are useful in the G0--M4 and G5--M3 range, respectively. Other lines sensitive to luminosity are listed by \citet{kh45}. They include the \ion{Fe}{i}~8327\AA\ line and the blend (mainly \ion{Fe}{i}) at 8514\AA. The behaviour of these lines was calibrated in early-M stars, and their behaviour at later types has not been calibrated. In any case, due to the growing strength of TiO bandheads, they cannot be used for types later than M4. It is so necessary to derive a spectral type first and then compare to stars of the same type to obtain a luminosity class.

For spectral types later than $\sim$M4, the measurement of most features becomes very difficult, as the spectra become dominated by TiO bands. 
As a consequence, the luminosity classification must be achieved by considering the relative strength of the whole metallic spectrum between $\sim$8660\AA\ and 8860\AA. This includes, apart from \ion{Ca}{ii}~8862\AA, the complex formed by \ion{Fe}{i}~8675\AA, \ion{Ti}{i}~8683\AA\ and \ion{Fe}{i}~8689\AA, the 8735\AA\ blend, the \ion{Mg}{i} triplet at 8807\AA\ and \ion{Fe}{i}~8824\AA\ \citep{allerkeenan51}. The luminosity class of late M stars with noisy spectra (where the strength of weak metallic lines is difficult to asses) cannot be considered very accurate. Note also that there are no MK standards for supergiants later than M5 \citep[cf.][]{solf}.

In view of the difficulties in classifying late-M stars, we have resorted to some approximate classifications. Some stars have strong TiO bandheads, indicative of spectral type M7 (i.e., $TiO\approx0.5$), but show relatively strong metallic lines and no sign of the VO 8624\AA\ bandhead. We have classified these objects as M6--7. Many other stars show evidence for the VO bandhead, and are therefore M7 or later, but their spectra are rather noisy (likely because the airglow is more difficult to remove due to much lower count rates inside the huge TiO features). In some cases, the metallic lines are quite strong, clearly indicating that the star is not a giant, but we cannot be sure that they are sufficiently strong to be a supergiant. In such cases, we have used the I--II classification, which is meant to indicate that the star is likely a very late supergiant, but may also be a bright giant.

\subsection{Radial velocities}
\label{radvels}

Radial velocities were calculated using the same algorithm described in \citetalias{neg11}. In that paper, we calculated cross-correlations against the spectra in the library of \citet{cenarro01}. In this work, however, we cross-correlate our spectra against a set of model spectra from the POLLUX database \citep{palacios10}. This change has been motivated by two reasons: firstly, the higher resolution available for the synthetic spectra makes them more suitable for the dataset we use here (especially, for the spectra obtained with ISIS, with $R>15\,000$). Secondly, the radial velocity corrections of  \citet{cenarro01} are not always accurate. Many luminous red stars -- particularly those of later spectral types -- display radial velocity variations, but most of the stars in the library of \citet{cenarro01} were observed only once and then corrected with a shift reflecting their average radial velocity. As a consequence, the use of these spectra as templates introduces additional dispersion in radial velocities. By correlating all the spectra in the library against the POLLUX synthetic data, we have found that this dispersion can be as high as $8\:{\rm km}\,{\rm s}^{-1}$ for some given stars.

The synthetic spectra were degraded to the instrumental resolution of our spectra. To reduce the uncertainty in radial velocity due to the choice of a template, every spectrum was calibrated against its closest match in spectral type, following the procedure described in \citetalias{neg11}. The internal accuracy of our measurements can be gauged from the several stars that were observed two or three times with AF2. The differences between the spectra were measured and the dispersion in $v_{\mathrm{rad}}$ was found to be $\sigma(v)=2.8\:{\rm km}\,{\rm s}^{-1}$. This value is surprisingly good when compared to the spectral resolution, but the fact that AF2 is a bench mounted spectrograph results in very high stability.

Our two sets of spectra (from AF2 and ISIS) were observed along with stars of known radial velocity, allowing us to derive absolute
errors for our measurements. The AF2 data include 3 supergiants already
observed in \citetalias{davies07}. The average difference between our values and those of \citetalias{davies07} is $\Delta v_{\mathrm{rad}}=-1\pm4\:{\rm km}\,{\rm s}^{-1}$, indicating that there are no systematic differences.

 For the ISIS dataset, 5 red giants with highly accurate $v_{\mathrm{rad}}$ measurements in \citet{mermilliod} were observed along with the science targets. Comparing our measured $v_{\mathrm{rad}}$ to those of \citet{mermilliod}, we find a difference $\Delta v_{\mathrm{rad}}=-1.2\pm0.3\:{\rm km}\,{\rm s}^{-1}$. Such difference is compatible with the highest accuracy in radial velocity determination achievable with ISIS, according to its manual.

Finally, we can compare the radial velocities in our two datasets.
The ISIS and AF2 datasets have 11 stars in common. Once the systematics have been removed, their velocities agree very well, with $\Delta v_{\mathrm{rad}}=1\pm5\:{\rm km}\,{\rm s}^{-1}$. In view of all this, we estimate that our radial velocities have a conservative total uncertainty of $\pm5\:{\rm km}\,{\rm s}^{-1}$.

The radial velocities measured were transformed into the Local Standard of Rest (LSR) reference system using IRAF's {\it rvcorrect} package. In this direction ($l\approx24\degr-30\degr$), the Galactic rotation curve allows the separation of stars located in different Galactic arms. The distribution of molecular clouds along the line of sight \citep{rathborne09} shows a small group of objects with $v_{\rm LSR}$\,$\approx+25\:{\rm km}\,{\rm s}^{-1}$, perhaps identifiable as the Sagittarius Arm (but see below). Then we find a number of sources with velocities in the $\sim 40-65\:{\rm km}\,{\rm s}^{-1}$ range, generally interpreted as a first crossing of the Scutum-Crux arm. There is a clear gap for $v_{\rm LSR}$\,$\approx70-90\:{\rm km}\,{\rm s}^{-1}$ (except for the $l\approx$\,$28\degr-30\degr$ range), after which we find a high number of sources with velocities between $90$ and $115\:{\rm km}\,{\rm s}^{-1}$. \citetalias{davies07} determine an average systemic velocity $v_{{\rm LSR}}$\,$=110\:{\rm km}\,{\rm s}^{-1}$ for Ste~2. The distribution of \ion{H}{ii} regions \citep{bania} confirms the low number of tracers associated with the Sagittarius arm, but reflects a more homogeneous distribution in velocities between $40$ and $90\:{\rm km}\,{\rm s}^{-1}$, with again a high concentration in the $v_{\rm LSR}$\,$\approx90-115\:{\rm km}\,{\rm s}^{-1}$ range.

\section{Results}
\label{sec:res}

Once again, the selection criteria have been very effective. All the objects observed are late-type luminous stars, with one single exception, J18390349$-$0546386, which is a B-type supergiant\fnmsep\footnote{The 2MASS magnitudes of this object clearly correspond to a late-type star. The DENIS $J$ and $K$ magnitudes are compatible within the errors. The object is saturated in UKIDSS, but, with the higher spatial resolution, it is clearly resolved into two stars, separated by $\sim1\farcs8$. We believe that the $I$-band magnitude is likely dominated by the second star, as there is no evidence for a late-type component in our spectrum.}. The rest of the AF2 targets comprise 3 G-type supergiants, a number of K-type stars with luminosity classes II and higher and M-type giants and supergiants. Early M-type stars have luminosity classes II and higher, while for spectral types later than M3, we also detect luminosity class III objects. This distribution of spectral types seems to confirm the hypothesis advanced in \citetalias{neg11} that our $Q_{{\rm IR}}$ cut selects stars with very extended atmospheres. Table~\ref{tab:results} presents the derived spectral types for all our targets and the colour excesses $E(J-K_{{\rm S}})$, calculated using the calibration of \citet{levesque}. 

\begin{table*}
\caption{List of AF2 targets with usable spectra with their main properties.\label{tab:results}}
\begin{tabular}{lcccccc}
\centering
ID&$J$&$H$&$K_{{\rm S}}$&$E(J-K_{{\rm S}})$&$V_{{\rm LSR}}$&Spectral\\
&&&&&$({\rm km}\,{\rm s}^{-1})$&type\\
\hline
\hline
18382013$-$0543434$^{a}$&$  8.37\pm 0.02$&$  6.70\pm 0.04$&$  5.95\pm 0.01$&  1.35&  +59&   M1.5\,Iab\\
18383438$-$0527029&$  8.94\pm 0.05$&$  7.52\pm 0.02$&$  6.95\pm 0.01$&  0.70&  +67&     M7\,III\\
18384011$-$0547537&$  7.42\pm 0.01$&$  6.10\pm 0.03$&$  5.48\pm 0.01$&  0.64&  +59&      M7\,II\\
18384958$-$0557222&$  8.95\pm 0.01$&$  7.20\pm 0.03$&$  6.39\pm 0.02$&  1.27& $-$12&     M7\,III\\
18385040$-$0549183&$  9.81\pm 0.02$&$  7.78\pm 0.04$&$  6.86\pm 0.03$&  1.70&  +91&    M6\,III\\
18385144$-$0600228 (D12)&$  7.22\pm 0.04$&$  5.94\pm 0.02$&$  5.35\pm 0.01$&  0.62&  +35&    M7\,II-III\\
18385268$-$0544207&$  8.30\pm 0.01$&$  7.08\pm 0.02$&$  6.62\pm 0.01$&  0.62&  +43& M1.5\,Ib-II\\
18385336$-$0557444 (D38)&$  8.57\pm 0.01$&$  7.32\pm 0.01$&$  6.79\pm 0.03$&  0.61&  $-$4&    M4.5\,II\\
...&...&...&...&...&...&...\\
\hline
\end{tabular}
\begin{list}{}{}
\item[]$^{a}$ The $JHK_{{\rm S}}$ magnitudes are from 2MASS.  Spectral types are based solely on our spectra. The colour excess has been calculated from the observed $(J-K_{{\rm S}})$ colour and the tabulated intrinsic colour, after \citet{levesque}.
\end{list}
\end{table*}

Table~\ref{tab:resultsisis} presents the same data for the ISIS targets. There are eleven sources in common with the AF2 sample, and spectral types and $v_{{\rm LSR}}$ have been derived independently with the aim of gauging the effect of resolution on the procedures used. The differences in spectral type are minimal, with an occasional very small disagreement in the luminosity class. 

\begin{table*}
\caption{List of ISIS targets. Objects marked with a $\dagger$ were not primary targets and their identifications are based on slit reconstruction. Such objects felt in the slit by chance and do not necessarily fulfill the selection criteria.\label{tab:resultsisis}}
\begin{tabular}{lcccccc}
\centering

ID&$J$&$H$&$K_{{\rm S}}$&$E(J-K_{{\rm S}})$&$V_{{\rm LSR}}$&Spectral\\
&&&&&$({\rm km}\,{\rm s}^{-1})$&type\\
\hline
\hline
18365544$-$0643364$^{a}$&$  8.35\pm 0.02$&$  7.34\pm 0.02$&$  6.95\pm 0.03$&  0.51&  +48&	 K4$\,$II\\
18372648$-$0628397$^{\dagger}$&$  8.88\pm 0.03$&$  6.95\pm 0.02$&$  6.09\pm 0.02$&  1.73&  +68&	 M1$\,$Ib\\
18372699$-$0628444&$  7.12\pm 0.02$&$  5.90\pm 0.02$&$  5.37\pm 0.02$&  0.50&  +66&    M6--7$\,$II\\
18382442$-$0522521&$  8.14\pm 0.03$&$  7.00\pm 0.03$&$  6.57\pm 0.02$&  0.44& $-21$&M3.5$\,$II--III\\
18383565$-$0603043&$  7.78\pm 0.02$&$  6.64\pm 0.02$&$  6.18\pm 0.02$&  0.47&  +36&   M3.5$\,$III\\
18384011$-$0547537&$  7.42\pm 0.02$&$  6.10\pm 0.03$&$  5.48\pm 0.02$&  0.64&  +24&  M7$\,$II--III\\
18391838$-$0600383 (D6) &$  7.72\pm 0.02$&$  5.92\pm 0.03$&$  5.07\pm 0.02$&  1.51& +108&   M3.5$\,$Iab\\
18391952$-$0559194$^{\dagger}$ (D19)&$  8.28\pm 0.02$&$  6.58\pm 0.04$&$  5.80\pm 0.02$&  1.38& +111&	 M2$\,$Ia\\
18391961$-$0600408 (D2) &$  6.90\pm 0.02$&$  5.05\pm 0.02$&$  4.12\pm 0.27$&  1.49& +121&	M7.5$\,$I\\
18392506$-$0610036&$  7.76\pm 0.05$&$  6.53\pm 0.04$&$  6.05\pm 0.02$&  0.58&  +61&  M3$\,$II--III\\
18394635$-$0559473&$  7.05\pm 0.02$&$  5.34\pm 0.02$&$  4.51\pm 0.02$&  1.36& +110&	 M4$\,$Ib\\
18395282$-$0535172&$  6.81\pm 0.02$&$  5.41\pm 0.03$&$  4.75\pm 0.03$&  0.74&  +56&    M8$\,$I--II\\
18402741$-$0556336$^{\dagger}$&$ 12.09\pm 0.03$&$ 10.84\pm 0.02$&$ 10.38\pm 0.03$&  0.58&  +21&	 M3$\,$II\\
18402744$-$0556176&$  8.58\pm 0.02$&$  7.44\pm 0.02$&$  6.95\pm 0.02$&  0.50&  +53&	 M3$\,$II\\
18403179$-$0529522&$  7.88\pm 0.03$&$  6.55\pm 0.04$&$  5.98\pm 0.02$&  0.61&  +36&  M7$\,$II--III\\
18405021$-$0552520&$  7.66\pm 0.03$&$  6.49\pm 0.03$&$  6.00\pm 0.02$&  0.41&  +71&  M6$\,$II--III\\
18405351$-$0517488&$  8.42\pm 0.03$&$  7.20\pm 0.04$&$  6.74\pm 0.03$&  0.69&  +71&	K7$\,$Iab\\
18405360$-$0525004&$  7.53\pm 0.02$&$  6.12\pm 0.05$&$  5.46\pm 0.02$&  0.93& +113&	 M3$\,$Ib\\
18410764$-$0535045&$  8.16\pm 0.02$&$  7.12\pm 0.03$&$  6.67\pm 0.02$&  0.25& $-26$&    M6--7$\,$II\\
18410851$-$0555046&$  8.18\pm 0.02$&$  6.98\pm 0.03$&$  6.50\pm 0.02$&  0.51&  +71&    M4.5$\,$II\\
18412383$-$0526073&$  7.79\pm 0.02$&$  6.43\pm 0.03$&$  5.87\pm 0.02$&  0.90& +103&	M0$\,$Iab\\
18414143$-$0531378&$  8.23\pm 0.04$&$  7.23\pm 0.04$&$  6.81\pm 0.02$&  0.61&  +73&	K3$\,$Iab\\
18415112$-$0610503&$  7.99\pm 0.02$&$  6.91\pm 0.03$&$  6.48\pm 0.03$&  0.31&  +21&   M5$\,$Ib--II\\
18415256$-$0611381$^{\dagger}$&$ 11.83\pm 0.02$&$ 10.13\pm 0.02$&$  9.42\pm 0.02$&  1.78&  $-1$&	K0$\,$III\\
18421965$-$0545389$^{\dagger}$&$ 10.32\pm 0.02$&$  9.30\pm 0.03$&$  8.87\pm 0.03$&  0.44&  +49&	M0$\,$Iab\\
18422147$-$0545383&$  8.53\pm 0.02$&$  7.41\pm 0.05$&$  6.98\pm 0.02$&  0.37&  +34&   M4.5$\,$III\\
\hline
\end{tabular}
\begin{list}{}{}
\item[]$^{a}$ The $JHK_{{\rm S}}$ magnitudes are from 2MASS.  Spectral types are based solely on our spectra. The colour excess has been calculated from the observed $(J-K_{{\rm S}})$ colour and the tabulated intrinsic colour, after \citet{levesque}.
\end{list}
\end{table*}

The number of true supergiants (i.e, objects with spectral types indicative of high-mass stars\fnmsep\footnote{These would be G and K stars of luminosity classes Ia and Iab, and all M-type supergiants.}) detected is quite high. The AF2 sample includes more than 60 supergiants (72 if we assume that late-M stars classified I-II and with $v_{\rm LSR}\geq90\:{\rm km}\,{\rm s}^{-1}$ must be supergiants) and 11 others with Ib-II classification. The ISIS sample adds 6 more RSGs. Of these 78 RSGs, only five are members of Ste~2 known from previous works. As a comparison, before the discovery of the RSG clusters, the number of confirmed RSGs in the Galaxy was just a few hundred. 

\begin{table*}
\caption{New spectral types for members of Stephenson~2.\label{tab:ste2}}
\begin{tabular}{lcccccccccccccc}
ID &   \multicolumn{2}{c}{Co-ordinates (J2000)}  & $i$ &$J$ & $H$    & $K_{{\rm S}}$ & Spectral & Spectral&$v_{{\rm LSR}}$&$v_{{\rm LSR}}$ D07\\
   &   RA     &    Decln.  &   &&&& Type (CO)$^{b}$& Type&(km\,s$^{-1}$)&(km\,s$^{-1}$)\\
\hline
\hline
D2  & 18 39 19.6 &$-$06 00 40.8&12.66&6.90 &5.05 &4.18 & M3\,I&M7.5\,I&+121&+111\\
D6 & 18 39 18.4 &$-$06 00 38.4&12.65&7.72& 5.92 &5.06& M5\,I&M3.5\,Iab&+108&+107\\
D15 & 18 39 22.4 &$-$06 01 50.1&13.07&8.13& 6.35 &5.51 & M2\,I&M1.5\,Iab&+106&+112\\
D18 & 18 39 22.5 &$-$06 00 08.4&12.88&8.18& 6.45 &5.63 & M4\,I&M0.5\,Iab&+114&+111\\
D19 & 18 39 19.5 &$-$05 59 19.4&13.01&8.28& 6.58 &5.80 & M2\,I&M1\,Iab& +111&+107\\
\hline
\end{tabular}
\begin{list}{}{}
\item[]$^{a}$ $JHK_{{\rm S}}$ magnitudes are from 2MASS.  $i$ magnitudes from DENIS.
\item[]$^{b}$ Estimated from the strength of the CO $2.3$~$\mu$m bandhead by \citetalias{davies07}. 

\end{list}
\end{table*}

We find an unexpectedly high number of stars with velocities in excess of $v_{{\rm LSR}}=+120\:{\rm km}\,{\rm s}^{-1}$, i.e., higher than any value expected along this line of sight, with a few values $v_{{\rm LSR}}\geq+145\:{\rm km}\,{\rm s}^{-1}$. Such stars must have high peculiar velocities.

\begin{figure}
\resizebox{\columnwidth}{!}{
\includegraphics[angle=0,clip]{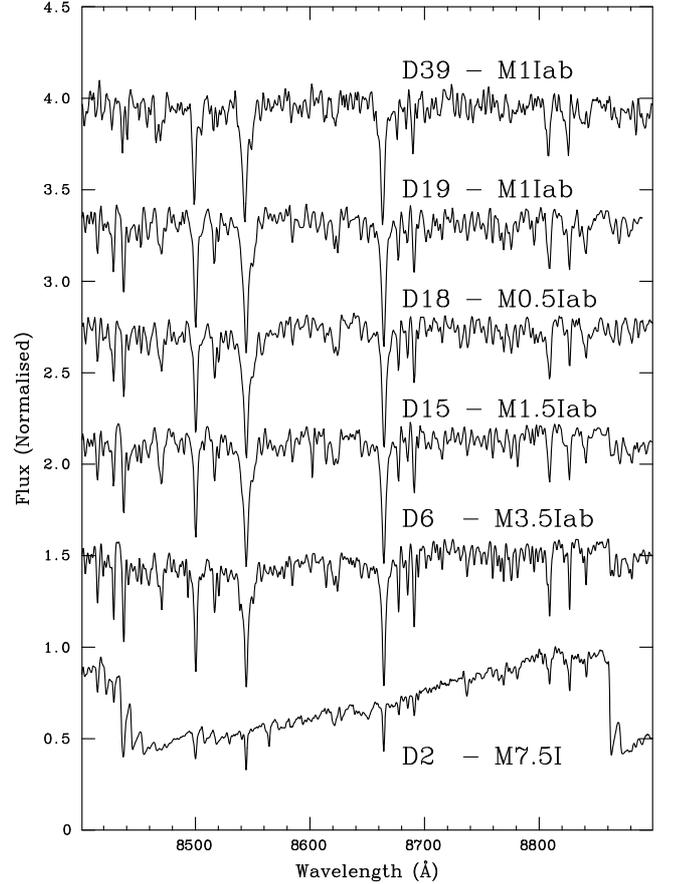}}
\caption{Spectra of five RSGs in Stephenson~2 and one possible runaway member (D39). The top four spectra are from AF2, while the bottom two are from ISIS, degraded to a similar resolution. Star D2 is a very late RSG, associated with an IRAS source and maser sources (compare to MY~Cep in Fig.~\ref{lates}). The spectrum of D39 has very low quality, as this is a faint target. \label{fig:rsgc2}}
\end{figure}

The presence of stars with very high peculiar velocities is demonstrated by several measurements of negative $v_{{\rm LSR}}$, which are not expected at all along this line of sight. Most of these objects are late M giants, the most extreme case being J18403943$-$0516043 (M4.5\,III) which displays $v_{{\rm LSR}}=-72\:{\rm km}\,{\rm s}^{-1}$. This object presents some excess in the far-IR (the 21-$\mu$m WISE point deviates considerably from a stellar model), but the {\it Spitzer}/GLIMPSE data do not indicate mid-IR excess. Therefore most of the $E(J-K_{{\rm S}})\approx0.65$ must be interstellar and this should be a relatively distant object, located no closer than the Sagittarius arm. Its peculiar velocity component is thus at least $-100\:{\rm km}\,{\rm s}^{-1}$. \citet{deguchib} found several maser sources with similar velocities at Galactic longitudes $l=20\degr-40\degr$, and interpreted them as members of streaming motion groups. Stars with $v_{{\rm LSR}}$\,$\ga120\:{\rm km}\,{\rm s}^{-1}$, on the other hand, might be following orbits profoundly disturbed by the Galactic Bar potential. Most of them have moderately high colour excess, suggesting that they are distant objects. As these objects might be signposts of systematic deviations from the Galactic rotation curve, the accuracy of kinematic distances along this line of sight may well be rather lower than generally assumed.

\subsection{Selection biases}

To assess the statistical value of our results, we need to evaluate the importance of selection biases in the sample finally observed. The most obvious source of bias is the fact that we are observing highly-reddened sources (almost) in the optical range. The choice of the observation range is determined by the high multiplexing capability of optical spectrographs, but also by the much better ability to obtain spectral classifications for red luminous stars in the 8300--8900\AA\ range than in the $K$ band, where our targets are intrinsically much brighter and extinction is much lower.

Because of this, it is encouraging to find that a very substantial fraction of the objects passing the 2MASS cuts, have a counterpart in the $I$ band (note that the $i$ or $I$ bands fall on the short wavelength side of our spectral range). In the main target area, 73\% of the 2MASS candidates have an $I$-band counterpart with $I<15.1$, and where so eligible for observing. In the control zone, the fraction was 68\% (though a few targets with fainter counterparts were also observed). Therefore the fraction of candidates too faint to be observed with our instrumentation was $\sim30\%$. 

Finally, it is worthwhile mentioning that there is no clear correlation between brightness in the $I$ band and spectral type. Even though a significant fraction of the catalogued members of Ste~2 have $I$ magnitudes that would have put them close of the faint end of our ``intermediate-brightness'' configuration, we find a very significant number of supergiants in the ``bright'' configuration. Many of them are clearly foreground to Ste~2, but there are a few RSGs with $v_{{\rm LSR}}>90\:{\rm km}\,{\rm s}^{-1}$ and rather high reddening that still manage to have $I<11.1$. In view of this, our ``bright'' cut at $I<10$ may perhaps seem too restrictive. In any case, less than 1\% of the 2MASS candidates were rejected by this magnitude cut.

The $K_{{\rm S}}=7.0$ cut was chosen as $\sim$90\% of the confirmed members of Ste~2 are brighter than $K_{{\rm S}}=6.2$. If a given star is fainter than the cluster members because of higher extinction, one would expect its $I$ magnitude to be much fainter than those of cluster members, making it an unpractical target for this survey. If, on the other hand, a star is fainter because of intrinsic lower luminosity, it is likely to represent an older or more distant population. Setting the limit at $K_{{\rm S}}=7.0$ seemed a sensible approach to account for these possibilities, while still keeping a reasonably small sample.

Simply moving the magnitude cut to $K_{{\rm S}}=7.5$ increases the total sample (objects observable in $I$ plus those not accessible) from slightly below 600 to more than 900 stars. For this increased sample, the fraction of candidates with a detection in $I$ and the fraction of candidates with $10\leq I<15.1$ are not significantly different from the  $K_{{\rm S}}=7.0$ sample. Analysis of the 2MASS CMDs and $Q/K_{{\rm S}}$ diagram, however, shows that for $K_{{\rm S}}\ga8$ field contamination becomes so important that a substantial number of objects fulfilling our colour criteria could be foreground F-type stars.

\subsection{Stephenson~2}

Care was taken to include some members of Ste~2, with radial velocities measured by \citetalias{davies07} to be used as calibration. Unfortunately, it is very difficult to place AF2 fibres on members of compact clusters, because they collide.  Three stars were observed with AF2, one of them (D19) included in two configurations. D19 and two additional stars were observed with ISIS. The list of Ste~2 stars is given in Table~\ref{tab:ste2}, together with the derived spectral types and a comparison of their properties in \citetalias{davies07}, including the spectral type estimation based on the strength of the CO $2.3$~$\mu$m bandhead. The spectra are displayed in Fig.~\ref{fig:rsgc2}.

We observed two stars in common with \citetalias{neg11}. The spectral type of D15 is M1.5\,Iab, as in the previous paper. 
The spectral type of D18, M0.5\,Iab, is slightly different from that found in \citetalias{neg11} (M1.5\,Iab), though we must note that differences in this spectral range are very subtle. This star was classified as M4\,I by \citet{davies07}. Another star presenting a large difference in spectral type is D2, which \citetalias{davies07} give as M3\,I, but appears very late here (we give M7.5\,I). This new classification, when considered together with its identification with the mid-IR sources IRAS~18366$-$0603 and MSX6C~G026.2038$-$00.0574, strongly points to this star (the brightest confirmed cluster member in $K_{{\rm S}}$) being surrounded by a dust envelope. \citet{deguchi} find both SiO and H$_{2}$O masers associated with this object (see later).

Our sample includes a few candidates from \citetalias{davies07} that were considered non-members or did not have their radial velocities measured. 
The first one is J18385144$-$0600228 (D12). Its moderate luminosity (M7\,II-III) and its $v_{{\rm LSR}}=+35\:{\rm km}\,{\rm s}^{-1}$ rule out membership. Its low reddening $E(J-K_{{\rm s}})\approx0.60$ is typical of the first crossing of the Scutum-Crux arm, at $\sim3.5$~kpc. Likewise, J18385336$-$0557444 (D38) is a bright giant (M4.5\,II) with $v_{{\rm LSR}}=-4\:{\rm km}\,{\rm s}^{-1}$. Again, its low reddening $E(J-K_{{\rm s}})\approx0.60$ confirms that it is a foreground star.

J18391636$-$0603149 (D28) was considered a foreground giant by \citetalias{davies07} because of the low EW of the $K$-band CO bandhead. Our spectrum (see Fig.~\ref{fig:early}), however, shows that it is a supergiant. The CO bandhead is weak because of its early spectral type, G5\,Iab. This star is projected very close to the central cluster concentration. Its $v_{{\rm LSR}}=+27\:{\rm km}\,{\rm s}^{-1}$ rules out, in principle, a connection to the cluster. Its $E(J-K_{{\rm S}})=1.0$ is lower than that of cluster members and the 8620\AA\ DIB is rather weak in its spectrum. This object is thus foreground to the cluster.

Finally, J18390701$-$0605043 (D39) is an M1\,Iab supergiant. Its $v_{{\rm LSR}}=+63\:{\rm km}\,{\rm s}^{-1}$ is very different from those of cluster members, but its $E(J-K_{{\rm S}})=1.79$ is typical for cluster members\fnmsep\footnote{The median value for the cluster is $E(J-K_{{\rm S}})=1.76$, calculated for the 26 likely members selected by \citetalias{davies07}, using the same calibrations as in that work.}. This star lies between D1 and D5, in the Southern concentration that \citet{deguchi} call Ste2~SW. Its spectral type is also typical of cluster members. If the star is foreground, it must be rather less luminous than its luminosity class indicates. Because of this, we cannot rule out the possibility that this is a cluster member with peculiar velocity.

Conversely, we find several stars outside the region studied by \citetalias{davies07} that are very likely connected to the cluster. For example, J18393774$-$0556183 is an M0\,Ia supergiant with  $E(J-K_{{\rm S}})=1.39$ and $v_{{\rm LSR}}=+98\:{\rm km}\,{\rm s}^{-1}$ only $\sim7\arcmin$ to the NE of the centre of Ste~2. Two arcmin further to the North, we find J18392977$-$0553032 (StRS 236), an M3.5\,Iab supergiant with $E(J-K_{{\rm S}})=1.48$ and $v_{{\rm LSR}}=+114\:{\rm km}\,{\rm s}^{-1}$. Slightly to the South, and just outside the region scanned by \citetalias{davies07}, we find J18394635$-$0559473 (IRAS~18370$-$0602 = MSX6C~G026.2678$-$00.1492), M4.5\,I, already classified as a candidate RSG by \citet{rawl00}. \citet{deguchi} excluded an association of this object with Ste~2 due to its spatial distance to the cluster, but the measured radial velocity $v_{{\rm LSR}}=+114\:{\rm km}\,{\rm s}^{-1}$ and $E(J-K_{{\rm S}})=1.36$ are comparable to those of cluster members.

Continuing in this area, J18393077$-$0551103 is an M1\,Iab supergiant $\sim$$11\arcmin$ North of the cluster core, with very high reddening $E(J-K_{{\rm S}})=1.98$ and $v_{{\rm LSR}}=+114\:{\rm km}\,{\rm s}^{-1}$.  J18393544$-$0548186 is a K7\,Iab supergiant $\sim$$14\arcmin$ away from Ste~2, with $E(J-K_{{\rm S}})=1.23$ and $v_{{\rm LSR}}=+106\:{\rm km}\,{\rm s}^{-1}$. Slightly further away, we find J18394562$-$0544057, M4.5\,Ib, with  $E(J-K_{{\rm S}})=1.08$ and  $v_{{\rm LSR}}=+113\:{\rm km}\,{\rm s}^{-1}$. All these objects seem to define an extension of the cluster towards the Northeast, without a clear divide. 

Interestingly, we also find two late-M giants with similar velocities, J18393324$-$0529435, M7.5\,III $(J-K_{{\rm S}}=3.2)$, and J18393145$-$0537096, M7\,III $(J-K_{{\rm S}}=2.8)$. Both are very red objects, and an important part of the obscuration must be interstellar\fnmsep\footnote{Theoretical magnitudes calculated from {\tt PHOENIX} models and calibrated against observational data suggest that all late-M giants have $(J-K)_{0}\sim1.3$ \citep{kuc05}. The calibration of \citet{groene04}, based on older models, gives $(J-K)_{0}=1.37$ for dust-free M7 stars. This reference also gives $(J-K)_{0}=1.97$ for dust-enshrouded M7 stars. None of the two stars is associated to an IRAS source, suggesting that they are not dust enshrouded. We note that IRAS~18368$-$0532 is unlikely to be associated to J18393324$-$0529435, but rather to J18393238$-$0530092, which is $\sim30\arcsec$ away and $\sim1.5$~mag brighter in all GLIMPSE bands.}. In view of this, they are likely to be at the same distance as Ste~2. We find several other examples of late M giants with LSR velocities $\geq90\:{\rm km}\,{\rm s}^{-1}$ and high reddenings at larger angular separations from Ste~2.

\subsection{Spatial distribution of RSGs}

\begin{figure*}
\resizebox{\textwidth}{!}{
\includegraphics[angle=90,clip]{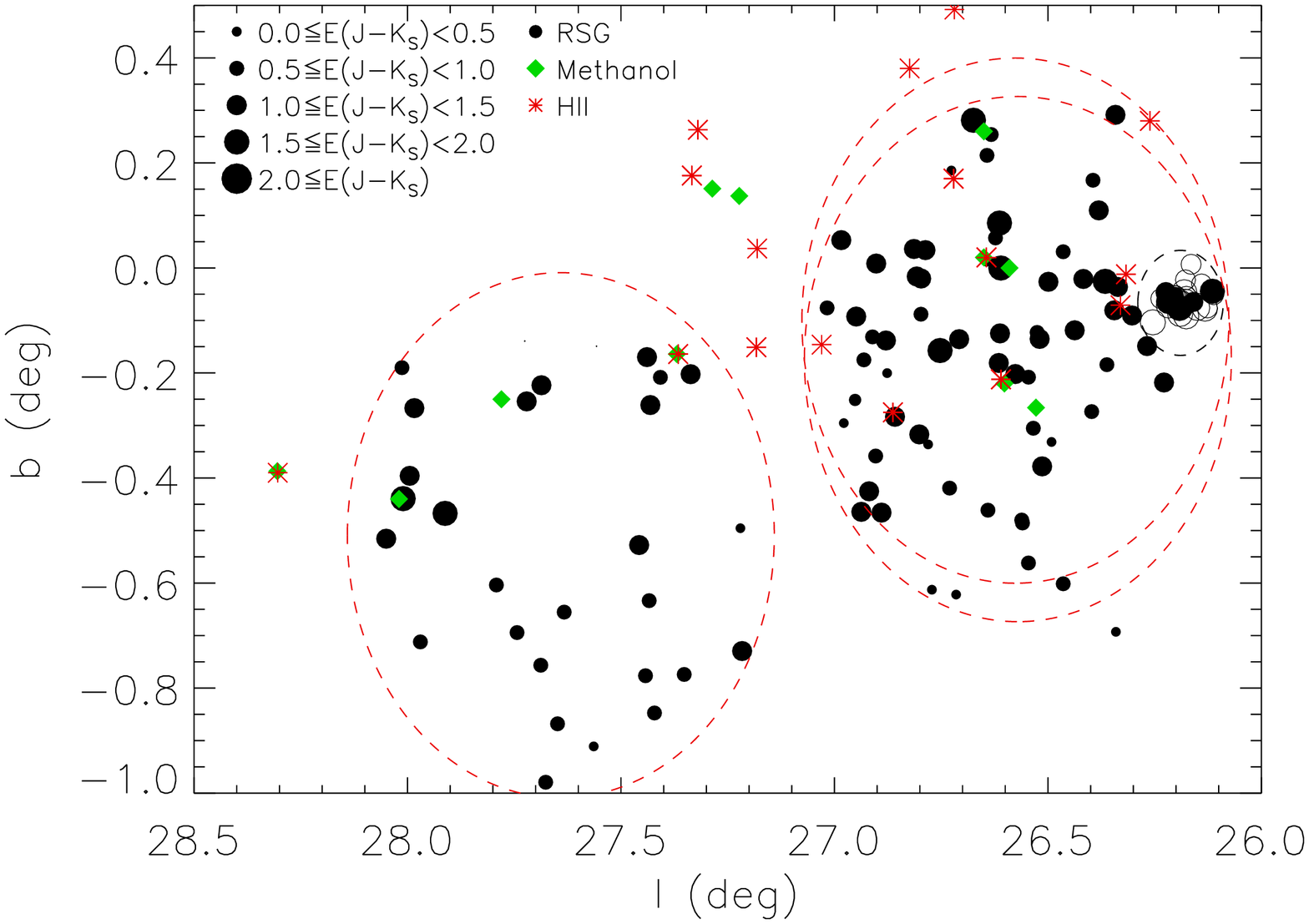}
\includegraphics[angle=90,clip]{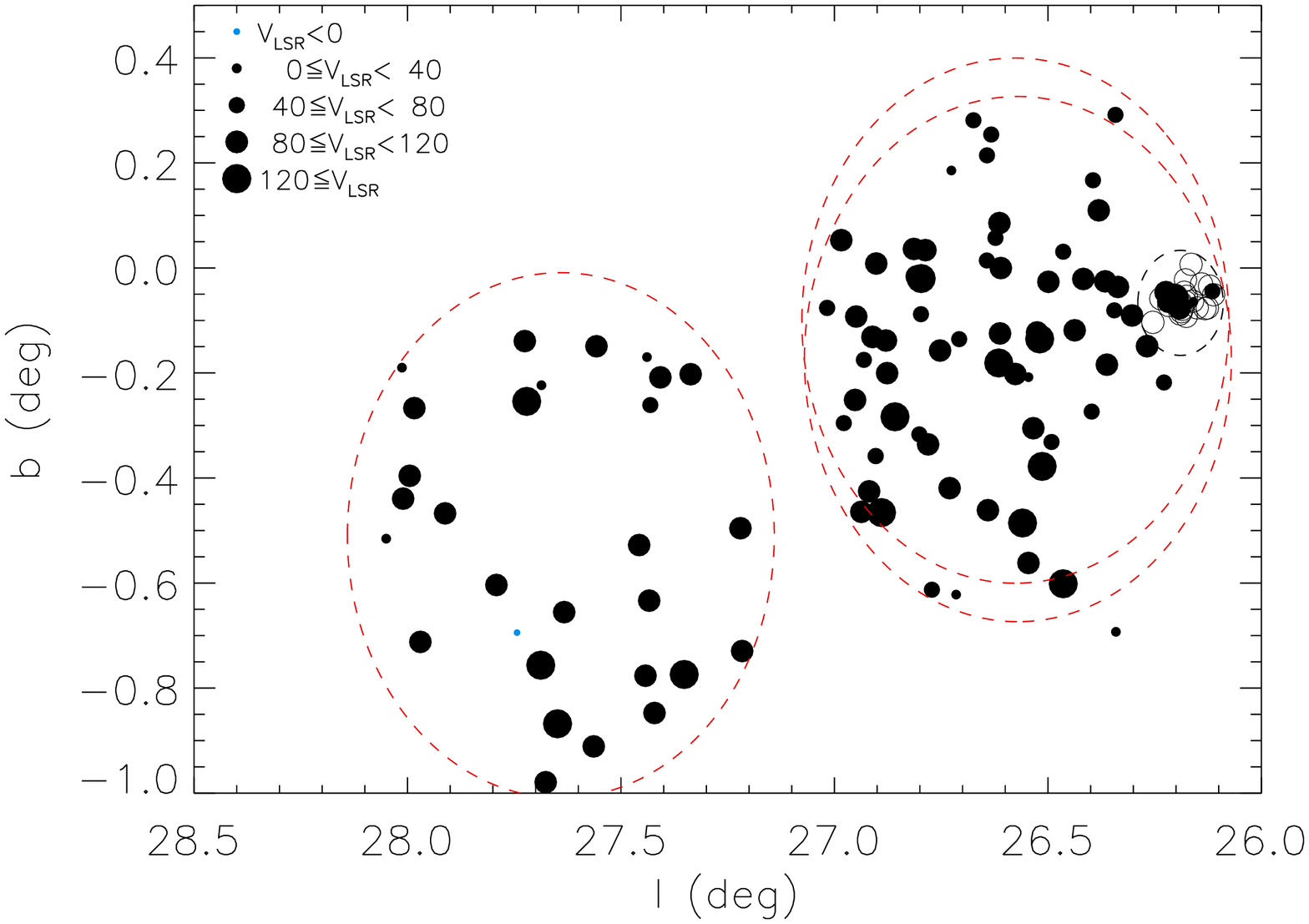}}
\caption{Spatial distribution of red supergiants detected in our survey. {\bf Left panel: Comparison to other spiral tracers}.  Black filled circles represent RSGs observed in our survey. Their size is proportional to the estimated $E(J-K_{{\rm S}})$, as shown in the legend. Open circles are other RSGs from \citetalias{davies07}. Diamonds are methanol masers (associated with star-forming regions) taken from \citet{green}. Asterisks represent H\,{\sc ii} regions observed in radio \citep{anderson11}. {\bf Right panel: Velocity distribution of RSGs.} The size of the symbols is proportional to their $v_{{\rm LSR}}$. The blue circle represents an RSG with $v_{{\rm LSR}}<0$ . In both panels, the nominal position of Ste~2 is indicated by a circle of radius $6\arcmin$. The large red ovals indicate the field of view of AF2 for the main target area (and the overlapping check exposure) and the control area (to the left).   \label{fig:distsg}  }
\end{figure*}

Figure~\ref{fig:distsg} shows the spatial distribution of the objects classified as RSGs, compared to the distribution of other spiral tracers, \ion{H}{ii} regions observed in radio from the catalogue of \citet{anderson11}, and methanol masers \citep[associated to young stellar objects;][]{green}. The two separate circular regions corresponding to the main target area (East) and the control zone (West) are indicated. Ste~2 lies on the Eastern side of the main target area. The number of RSGs in the circle including Ste~2 is obviously higher (about two thirds of the total number), but this area was observed with four different configurations (plus the extra ISIS targets), against only one configuration for the control region. In view of this, {\em we find no compelling evidence suggesting that RSGs are more numerous in the region surrounding Ste~2, outside the central concentration, than in the control region}.

\begin{figure*}
\resizebox{\textwidth}{!}{
\includegraphics[angle=90,clip]{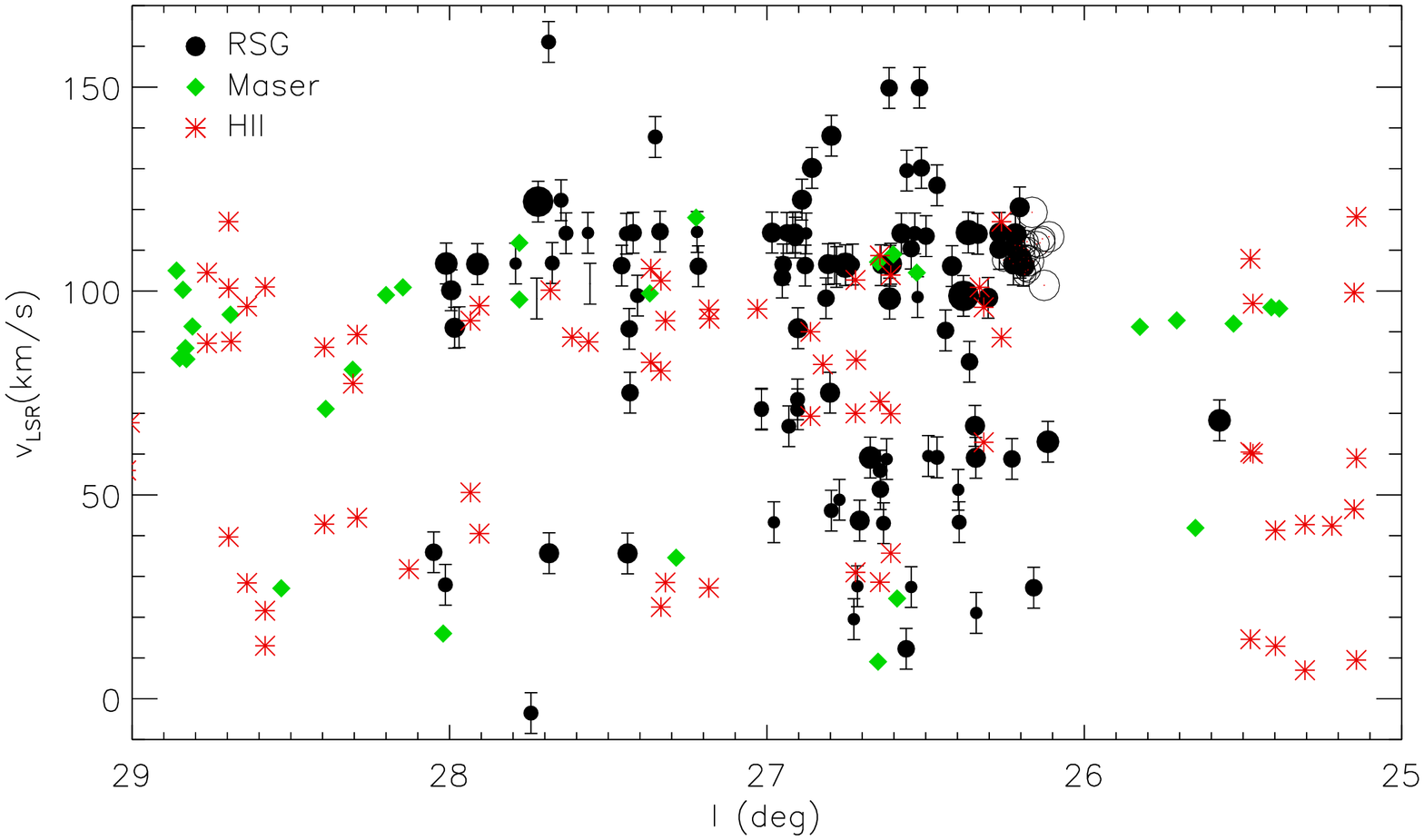}}
\caption{Diagram showing the ($v_{{\rm LSR}}$, $l$) distribution for different spiral tracers in the vicinity of Stephenson~2. Black circles represent red supergiants observed in our survey. Their size is proportional to the estimated $E(J-K_{{\rm S}})$, as in Fig.~\ref{fig:distsg}. Diamonds are methanol masers (associated with star-forming regions) taken from \citet{green}. Asterisks represent H\,{\sc ii} regions observed in radio \citep{anderson11}.  Cluster members from \citetalias{davies07} are shown as empty circles. 
\label{fig:lv}}
\end{figure*}

We only observed 71 stars in the control area, out of $\sim156$ possible targets, detecting 27 RSGs (38\%). In the main target area, however, we observed 162 stars out of 246 possible targets, detecting 45 RSGs (28\%). Therefore, even taking into account the fact that we cannot probe effectively the central region of  Ste~2, the fraction of RSGs in the control zone is not significantly different from that in the main target area. Moreover, most of the 27 RSGs detected in the control zone have $v_{{\rm LSR}}$ velocities $\geq90\:{\rm km}\,{\rm s}^{-1}$. This is more easily seen in Fig.~\ref{fig:lv}, which shows the distribution in $v_{{\rm LSR}}$ of all the RSGs detected in our survey compared to other spiral tracers. Over the whole $l=26\degr-28\degr$ range, there is a high concentration of sources with $v_{{\rm LSR}}\approx115\:{\rm km}\,{\rm s}^{-1}$. A few methanol masers have similar velocities, while \ion{H}{ii} regions prefer velocities slightly below $110\:{\rm km}\,{\rm s}^{-1}$. Many other RSGs, have velocities between 100 and $110\:{\rm km}\,{\rm s}^{-1}$. This concentration of sources delimits the terminal velocity in this direction. Different Galactic rotation curves imply terminal velocities betwen $v_{{\rm LSR}}=105$ and $120\:{\rm km}\,{\rm s}^{-1}$, depending on adopted values for the Sun's orbital velocity and distance to the Galactic centre \citep[e.g.,][]{clemens,levine08,reid09}.

We find many RSGs close to Ste~2 with $v_{{\rm LSR}}$ almost identical to cluster members, but a substantial population of RSGs with similar velocities extends to very large angular distances from the cluster. This result confirms the claim by \citet{garzon} of a strong over-density of RSGs towards $l=27\degr$, and demonstrates that this over-density exists over a much larger area than sampled by \citet{lopez}.

On the other hand, the most unexpected result of our survey is the large number of RSGs with velocities significantly smaller than the terminal velocity. We find $\sim$25 RSGs (not counting object classified Ib-II) with velocities $v_{{\rm LSR}}<90\:{\rm km}\,{\rm s}^{-1}$. A few of these objects have spectral types compatible with being (but not necessarily implying) intermediate-mass ($\sim$6--$8\,M_{\sun}$) stars, such as J18395686$-$0603482 (G8\,Ib) or J18410523$-$0525076 (K2\,Ib). Most of them, however, have spectral types suggestive of true supergiants (i.e., high-mass stars). Their number seems surprisingly high when compared to the census of RSGs known in the Galaxy or the Magellanic clouds, as they are simply the foreground population, likely to represent mostly the field population of the Scutum-Crux arm (see below).

\begin{figure*}
\resizebox{\textwidth}{!}{
\includegraphics[angle=90,clip, width=7cm]{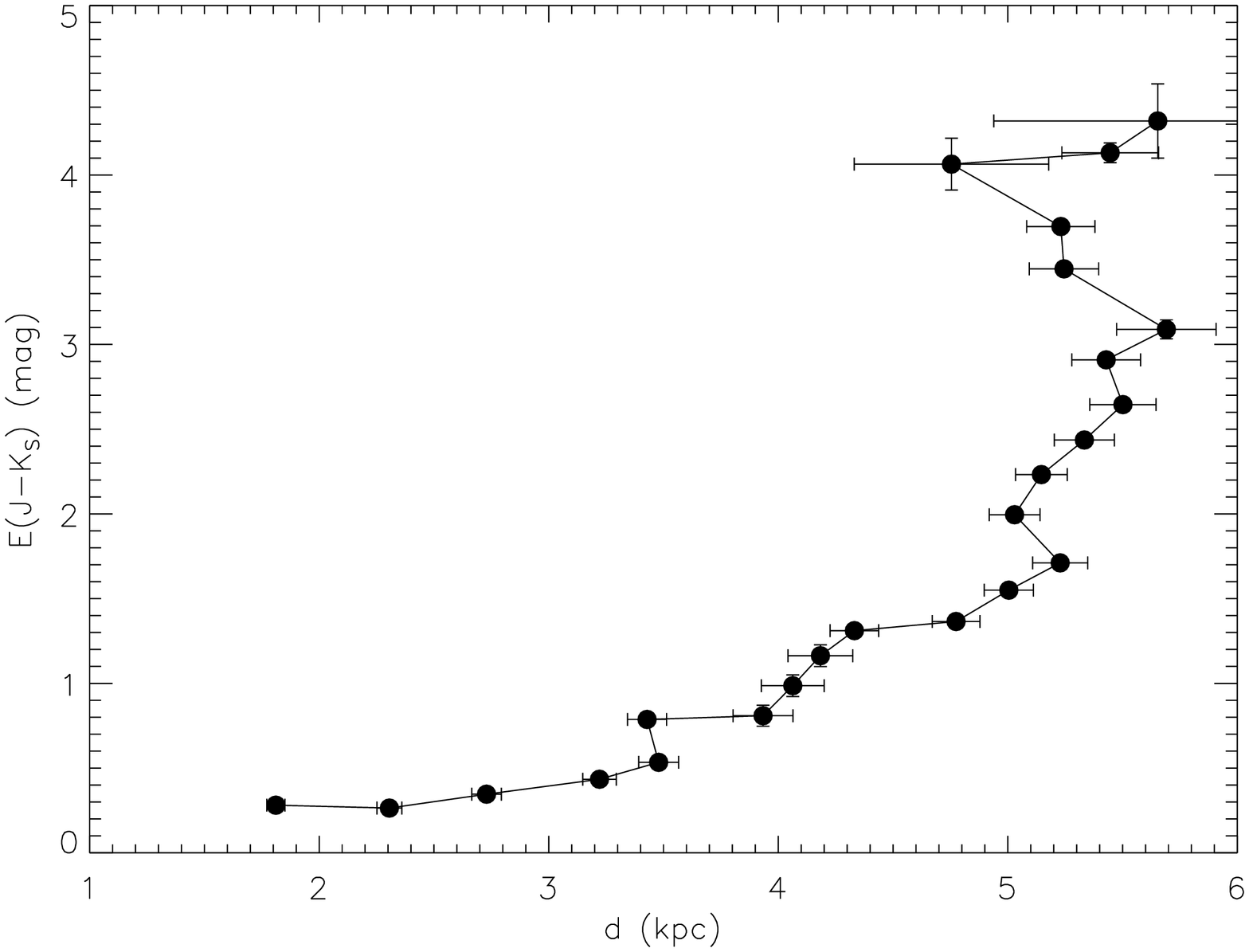}
\includegraphics[angle=90,clip, width=7cm]{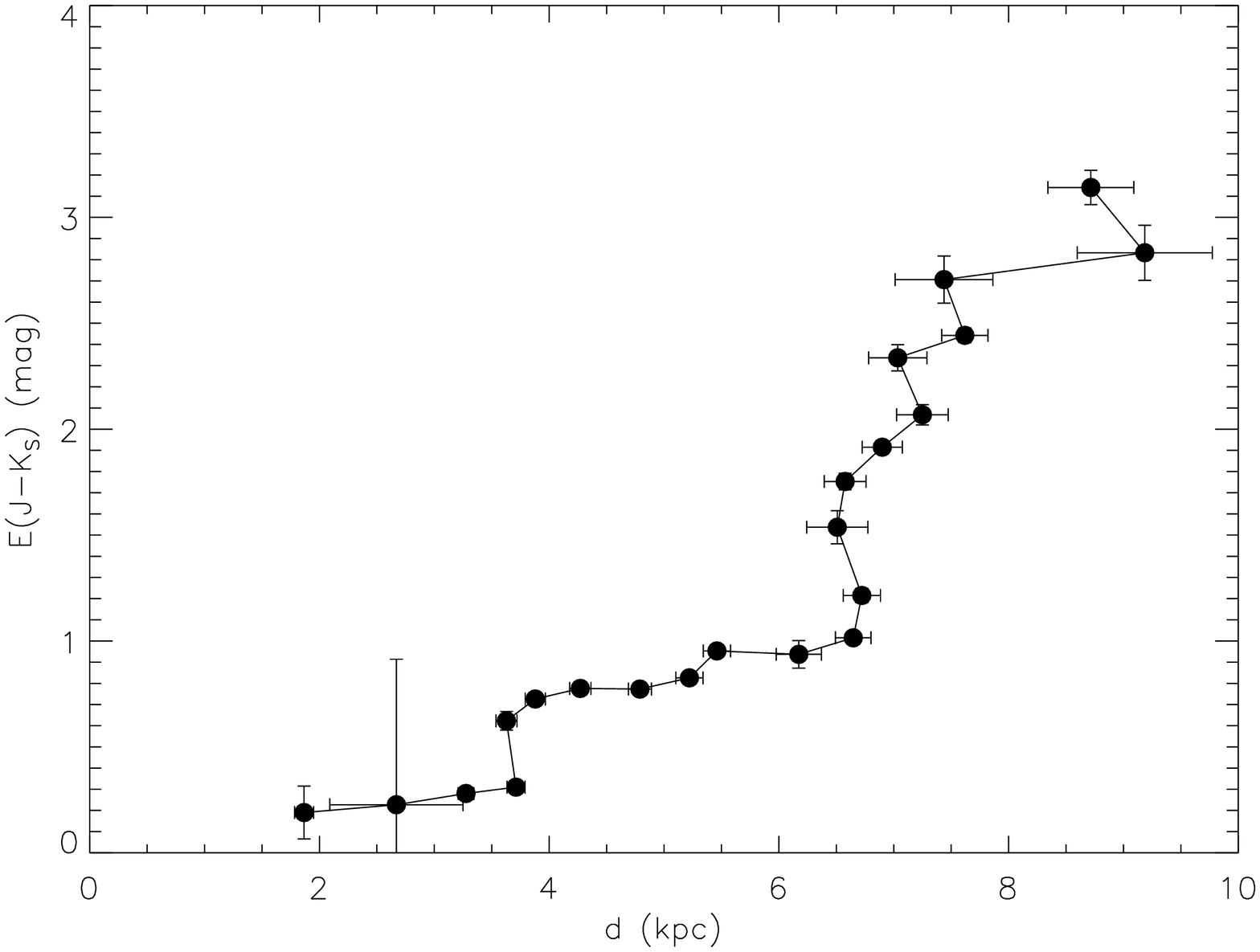}
\includegraphics[angle=90,clip, width=7cm]{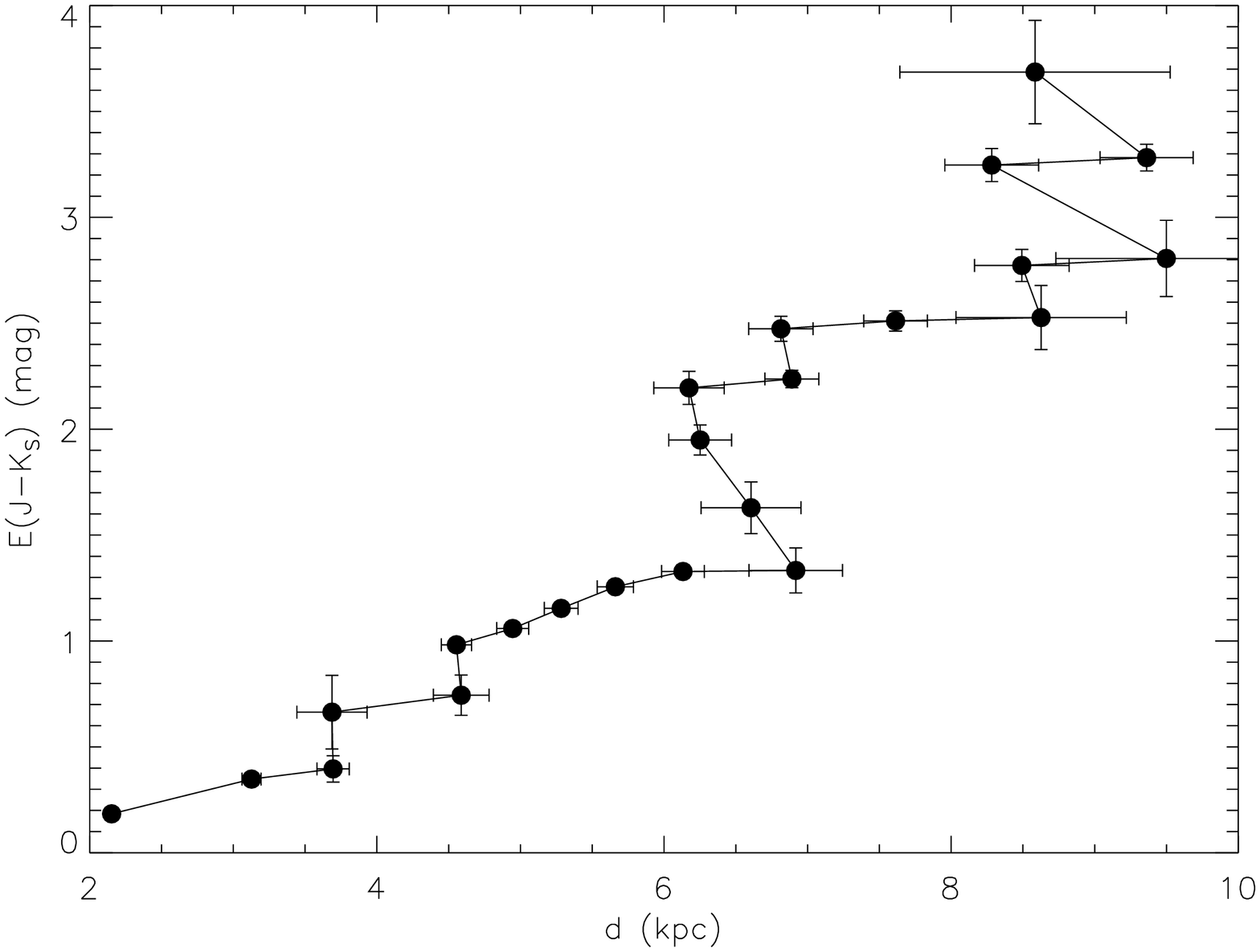}}
\caption{Run of the extinction in several directions covered by our survey. The data have been obtained by applying the technique of \citet{cabrera05} to the red clump giant population within $15\arcmin$ of the field centre. The left panel shows the results for the field centred on Stephenson~2. The middle panel represents a field centred on ($l=26\fdg5$, $b=-0\fdg5$). The right panel corresponds to a field centred on ($l=27\fdg5$, $b=-0\fdg5$).\label{fig:ext}}
\end{figure*}

\section{Discussion}
\label{sec:discu}

We have found high numbers of RSGs, with very different radial velocities and reddenings, spread around the galactic equator between longitudes $l=26\degr$ and $l=28\degr$. These objects must be tracing recent events of star formation. Here we try to understand their 3D spatial distribution by making use of the data available.

\subsection{The sightline to Stephenson~2}

Figure~\ref{fig:ext} shows the run of the extinction towards three directions covered by our survey, calculated using the technique of \citet{cabrera05} on UKIDSS data for circles of radius $15\arcmin$. This technique identifies the red clump giant population and estimates the average colour excess at a given distance modulus. The three figures are very similar to the corresponding diagrams in the directions to Alicante~8 \citep[$l=24\fdg6$;][]{neg10} and the RSGC3 association \citepalias[$l\approx29\fdg2$;][]{neg11}. Along all the sightlines, there is a relatively abrupt increase in the absorption around $d\approx3.5$~kpc, identifiable as the first crossing of the Scutum-Crux arm, where $E(J-K_{{\rm S}})$ climbs up from low values to $\approx0.8$. After this, the extinction grows gradually until $\sim$6~kpc, where it suddenly becomes so high as to render the technique unusable. The line of sight to Alicante~8 is slightly different, as the extinction remains relatively stable after 3.5~kpc and the sudden increase happens at $\sim$5~kpc, in front of the cluster.

The run of extinction to Ste~2 is also slightly different in one important aspect: the sudden rise in extinction happens slightly beyond 5~kpc. Taken at face value, the extinction curve places Ste~2 at $\approx5.3$~kpc. This value is compatible within the errors with previous determinations (\citealt{ortolani}, \citetalias{davies07}), though it lies near the low edge and forces the age value towards the high-end of the permitted range ($\sim20$~Myr). This distance value, though, should not be considered very accurate, as the extinction is determined over a $15\arcmin$ circle and there is clear evidence of very patchy absorption over the field. Maps of CO emission in this direction from the Galactic Ring Survey \citep{putaringonit} show very little emission below $v_{{\rm LSR}}$\,$=+90\:{\rm km}\,{\rm s}^{-1}$ and very strong emission for higher velocities.

All the directions show a very low amount of extinction associated with the Sagittarius arm, at $\sim$$1.5$--2~kpc. Our sample contains very few possible tracers of this arm, but this is fully expectable, as our colour cuts are chosen to leave out stars with little obscuration. For example, the open cluster NGC~6664 ($d\approx1.4$~kpc, $l=24\fdg0$; \citealt{mermi87,mermilliod}) contains several G--K\,II stars, all of which have $I<9$ and would thus have been rejected by our $I$-band brightness cut. 

In any case, spiral tracers for the Sagittarius Arm are scarce in this area. The maps of molecular clouds show a few objects with velocities $\approx$$+25\:{\rm km}\,{\rm s}^{-1}$ \citep{rathborne09}. The \ion{H}{ii} regions G026.103$-$00.069a\fnmsep\footnote{Radial velocities for \ion{H}{ii} regions are from \citet{anderson11} and references therein.} ($v_{{\rm LSR}}$\,$=+29.3\:{\rm km}\,{\rm s}^{-1}$) and G026.600$-$00.020a ($v_{{\rm LSR}}$\,$=+21.6\:{\rm km}\,{\rm s}^{-1}$) have velocities compatible with this arm, but the fact that they are not visible in H$\alpha$ casts doubt on the possibility that their distance is $<2$~kpc. Moreover, no clusters with ages $\leq20$~Myr and distances compatible with the Sagittarius arm are known. 

In our sample, there are a few bright giants with velocities in this range and relatively low reddenings. Good candidates to be located in the Sagittarius arm are J18395177$-$0549409 (M5\,II-III, $v_{{\rm LSR}}=+19\:{\rm km}\,{\rm s}^{-1}$), which has $E(J-K_{{\rm S}})\approx0.4$, or J18392533$-$0526088 (M2\,Ib-II, $v_{{\rm LSR}}$\,$=+20\:{\rm km}\,{\rm s}^{-1}$), with the same $E(J-K_{{\rm S}})$. The only supergiants that could be located in the Sagittarius arm are J18421727$-$0548553 (M5\,Ib, $v_{{\rm LSR}}=+28\:{\rm km}\,{\rm s}^{-1}$), with $E(J-K_{{\rm S}})=0.44$, and J18391636$-$0603149 (= D28;  G5\,Iab, $v_{{\rm LSR}}=+27\:{\rm km}\,{\rm s}^{-1}$)\fnmsep\footnote{In principle, the observed $E(J-K_{{\rm S}})\approx1.0$ suggests a much higher distance. We can estimate the luminosity using the calibration of the strength of the \ion{O}{i}~7774\AA\ by \citet{arellano}. We measure EW$=-0.6$\AA, corresponding to $M_{V}=-3.0$, according to this calibration. Using the intrinsic colour for a G5 supergiant, $M_{K}\approx-4.8$. This would imply a distance of only $\sim1.1$~kpc. Of course, the luminosity may be underestimated because the strength of the feature is decreasing for late-G spectral types (it has disappeared by K0), while the calibration has very large uncertainties (standard errors $\pm0.4$~mag). However, given these values, the source is likely to be in the Sagittarius arm. With a G5\,Iab spectral type, the star may be a Cepheid variable. If this is the case, its 2MASS colour might not correspond to the observed spectral type.}.

Towards $l=28-29\degr$, \citet{turner80} found a population of O-type stars and B supergiants at $d\approx3.5$~kpc and relatively low reddenings, $E(B-V)=1.1$--1.3 (similar to the colour excess of stars in the Sagittarius arm in the same direction). Likewise, towards $l=25\fdg3$, \citet{rcw} found OB stars with $E(B-V)\approx1.0$, and evidence for obscuring clouds at a distance $\sim3$~kpc. Several \ion{H}{ii} regions with $v_{{\rm LSR}}$\,$\approx+40\:{\rm km}\,{\rm s}^{-1}$ can be found in both directions (see Fig.~\ref{fig:lv}). We can identify this populations with the near side of the Scutum-Crux arm, just in front of the clouds causing absorption at $\sim$3.5~kpc.

In view of this, we would expect to see stars with $v_{{\rm LSR}}\approx40\:{\rm km}\,{\rm s}^{-1}$ and $E(J-K_{{\rm S}})\approx0.6$ tracing the near edge of the Scutum-Crux arm in our region. However, we find very few objects with such parameters between $l=26\degr$ and $l=28\degr$\fnmsep\footnote{On the Western side, only J18385268$-$0544207 (M1.5\,Ib--II,  $v_{{\rm LSR}}$$=+43\:{\rm km}\,{\rm s}^{-1}$), J18390034$-$0529144 (K4\,Ib--II,  $v_{{\rm LSR}}$$=+43\:{\rm km}\,{\rm s}^{-1}$) and perhaps J18385144$-$0600228 (M6\,I--II, $v_{{\rm LSR}}$$=+35\:{\rm km}\,{\rm s}^{-1}$) seem possible tracers of such structure. These three objects lie quite close together and very near the edge of the region surveyed (i.e., close to $l=26\degr$). Towards $l=28\degr$ we also find a few stars with $v_{{\rm LSR}}\approx+36\:{\rm km}\,{\rm s}^{-1}$, but much higher reddenings. Objects like J18423886$-$0446094, M1.5\,Iab and $E(J-K_{{\rm S}})\approx1.3$, or J18442147$-$0434460, K2\,Iab and $E(J-K_{{\rm S}})\approx1.0$, might be RSGs associated with high-obscuration regions in the Scutum-Crux arm.}. This low number of tracers is not due to our colour cuts, as our lower limit $(J-K_{{\rm S}})>1.3$ will only leave out stars with intrinsic colour  $(J-K_{{\rm S}})_{0}<0.7$, i.e, earlier than $\sim$K0, at this reddening. 

In contrast, between $l\approx26\degr$ and $27\degr$, we find RSGs with $v_{{\rm LSR}}=60$--$70\:{\rm km}\,{\rm s}^{-1}$ and moderately high colour excesses. They seem to be accompanied by a large number of \ion{H}{ii} regions\fnmsep\footnote{In the catalogue of \citet{anderson11}, we find G025.469$-$0.121b ($v_{{\rm LSR}}$=$60.1\:{\rm km}\,{\rm s}^{-1}$), G025.477+0.040a ($v_{{\rm LSR}}$=$60.5\:{\rm km}\,{\rm s}^{-1}$), G26.317$-$0.012b ($v_{{\rm LSR}}$=$62.9\:{\rm km}\,{\rm s}^{-1}$),  G026.600$-$00.020c ($v_{{\rm LSR}}$=$69\:{\rm km}\,{\rm s}^{-1}$), G026.600$-$00.106 ($v_{{\rm LSR}}$=$69\:{\rm km}\,{\rm s}^{-1}$), G26.610$-$0.212b ($v_{{\rm LSR}}$=$69.9\:{\rm km}\,{\rm s}^{-1}$), G26.644+0.020b ($v_{{\rm LSR}}$=$72.9\:{\rm km}\,{\rm s}^{-1}$), G26.721+0.170b ($v_{{\rm LSR}}$=7$0\:{\rm km}\,{\rm s}^{-1}$) and G26.863$-$0.275a
($v_{{\rm LSR}}$=$69.3\:{\rm km}\,{\rm s}^{-1}$).}. This is clearly the only structure visible in Fig.~\ref{fig:lv} in this $l$ range. Its interpretation is difficult. The dynamical distance for this velocity range is  $\ga$4~kpc, but the extinction does not seem to increase significantly at this distance. We could think that this structure is tracing the far side of the Scutum-Crux arm. Other possible interpretation would assign (some of) the \ion{H}{ii} regions to the 3~kpc arm, though the RSGs in this velocity range seem to have too low reddening to be that far. Finally, we could also speculate with the possibility that this feature is due to a strong systematic deviation from the Galactic rotation curve in this direction.

Finally, a few \ion{H}{ii} regions\fnmsep\footnote{We find G026.433+00.614  ($v_{{\rm LSR}}$$=89\pm5\:{\rm km}\,{\rm s}^{-1}$), G026.536+00.416 ($v_{{\rm LSR}}$$=87\pm5\:{\rm km}\,{\rm s}^{-1}$) and G026.261+0.280a ($v_{{\rm LSR}}=$$88.6\pm0.4\:{\rm km}\,{\rm s}^{-1}$).} in the neighbourhood of Ste~2 present velocities around $v_{{\rm LSR}}$\,$=+90\:{\rm km}\,{\rm s}^{-1}$. There are several supergiants with $v_{{\rm LSR}}$ close to these values\fnmsep\footnote{For instance, J18395861$-$0549549 (G5\,Ia),  J18402277$-$0521377 (M0\,Ia), J18421794$-$0513098 (M7\,I--II), J18432089$-$0431281 (K3\,Iab) or J18433896$-$0510524 (M7\,I--II)}. All these objects have $E(J-K_{{\rm S}})\ga1.0$, suggestive of high distances. 

There is thus some evidence for a small difference between the sightline to Ste~2 and neighbouring ones, with a clear lack of spiral tracers in the $v_{{\rm LSR}}$$\approx40-60\:{\rm km}\,{\rm s}^{-1}$ range, in sharp contrast to the $l=24\degr-26\degr$ range, where \ion{H}{ii} regions and molecular clouds are very numerous, and the $l$\,$=28\degr-30\degr$ range, where they are found in moderate numbers. The relative distributions of these tracers may explain the very different reddenings to the RSG clusters in the Scutum Complex. Conversely, just in front of Ste~2, there is a fair number of tracers around $v_{{\rm LSR}}$\,$\approx70\:{\rm km}\,{\rm s}^{-1}$, not found in the neighbouring regions (see Fig.~\ref{fig:lv} and \citealt{rathborne09}).

\subsection{Nature and extent of Stephenson~2}

\citetalias{davies07} claimed 26 likely members of Ste~2 by selecting stars with $100\:{\rm km}\,{\rm s}^{-1}\leq v_{{\rm LSR}}\leq120\:{\rm km}\,{\rm s}^{-1}$. This choice resulted in an average radial velocity for the cluster of $v_{{\rm LSR}}=+109.3\pm0.7\:{\rm km}\,{\rm s}^{-1}$. The velocity range chosen is somewhat arbitrary. Indeed, D23, with $v_{{\rm LSR}}=+119.3\:{\rm km}\,{\rm s}^{-1}$ is a clear outlier. Star D17, with $v_{{\rm LSR}}=+101.4\:{\rm km}\,{\rm s}^{-1}$ is also far away from the cluster average, and has a CO bandhead EW well below that of all other members. The other 24 objects, however, form a very compact group in velocity space with  $104\:{\rm km}\,{\rm s}^{-1}\leq v_{{\rm LSR}}\leq113\:{\rm km}\,{\rm s}^{-1}$.

Is this range compatible with a single cluster? \citet{mermilliod} measured accurate radial velocities for several RSGs in Northern clusters, finding aperiodic variability in all of them, with variable amplitude in function of time. The full amplitudes observed ranged from $\la5$ to $10\:{\rm km}\,{\rm s}^{-1}$. In view of this result, it is fully expectable to find a given RSG with $v_{{\rm rad}}$ up to $5\:{\rm km}\,{\rm s}^{-1}$ away from the cluster average, even if we do not consider the cluster's dynamical velocity dispersion. Moreover, all the RSGs surveyed by \citet{mermilliod} have moderate luminosities and rather early spectral types. It is interesting to note that the two stars with latest spectral types, V589~Cas (M3\,Iab--Ib) and RS~Per \citep[M4\,Iab;][]{humph70}, display the largest amplitudes. It is possible to envisage later-type M supergiants, many of which are dust-enshrouded, as the high-mass analogues of extreme AGB stars, subject to heavy pulsation. Objects like VX~Sgr (M4e--M10e\,Ia) or S~Per (M3\,Iae) display changes in spectral type \citep{wing} and irregular photometric variations with full amplitudes in excess of 5~mag in the $V$ band \citep{kiss06}. Therefore even larger variations in $v_{{\rm rad}}$ are not discarded.

Two very bright stars associated with IRAS sources, D1 and D4, were not considered members by \citetalias{davies07}\fnmsep\footnote{Though \citetalias{davies07} argue that, given its extreme IR luminosity, D1 could be a cluster member experimenting some sort of expansion phase associated to extreme mass loss.}, as they both had  $v_{{\rm LSR}}\,\approx\,+95\:{\rm km}\,{\rm s}^{-1}$. On the other hand, \citet{deguchi} observed masers associated to five sources in the area. For D2, they measured velocities $v_{{\rm LSR}}$$\approx+102.7\:{\rm km}\,{\rm s}^{-1}$ and  $v_{{\rm LSR}}$$\approx+106.8\:{\rm km}\,{\rm s}^{-1}$ in the SiO $J = 1$--0 $\nu=1$ and H$_{2}$O $6_{16}$--$5_{23}$ transitions, respectively. For D49, they measured $v_{{\rm LSR}}$$\approx+102.7\:{\rm km}\,{\rm s}^{-1}$ in the SiO transition. Finally, for D1, they measured $v_{{\rm LSR}}$$\approx+92.7\:{\rm km}\,{\rm s}^{-1}$ and $v_{{\rm LSR}}$$=+95.6\:{\rm km}\,{\rm s}^{-1}$. The small variations with respect to the data in \citetalias{davies07} are likely attributable to the very different layers being measured. \citet{deguchi} argue that the group of sources to the Southwest of the cluster (D1, D5 and D49, plus another source outside the area surveyed by \citetalias{davies07}, which they call St2--26, and perhaps some others) are too far away from the cluster core to be physically related and must form a different cluster, which they name Ste~2~SW, with some (but non-significant) evidence for lower $v_{{\rm LSR}}$. They also find $v_{{\rm LSR}}$$=+86.7\:{\rm km}\,{\rm s}^{-1}$ in the  H$_{2}$O $6_{16}$--$5_{23}$ transition (their only detection) of D9. This is very different from the $v_{{\rm LSR}}$$=+111.8\:{\rm km}\,{\rm s}^{-1}$ measured by \citetalias{davies07}.

\begin{figure}
\resizebox{\columnwidth}{!}{
\includegraphics[angle=-90,clip]{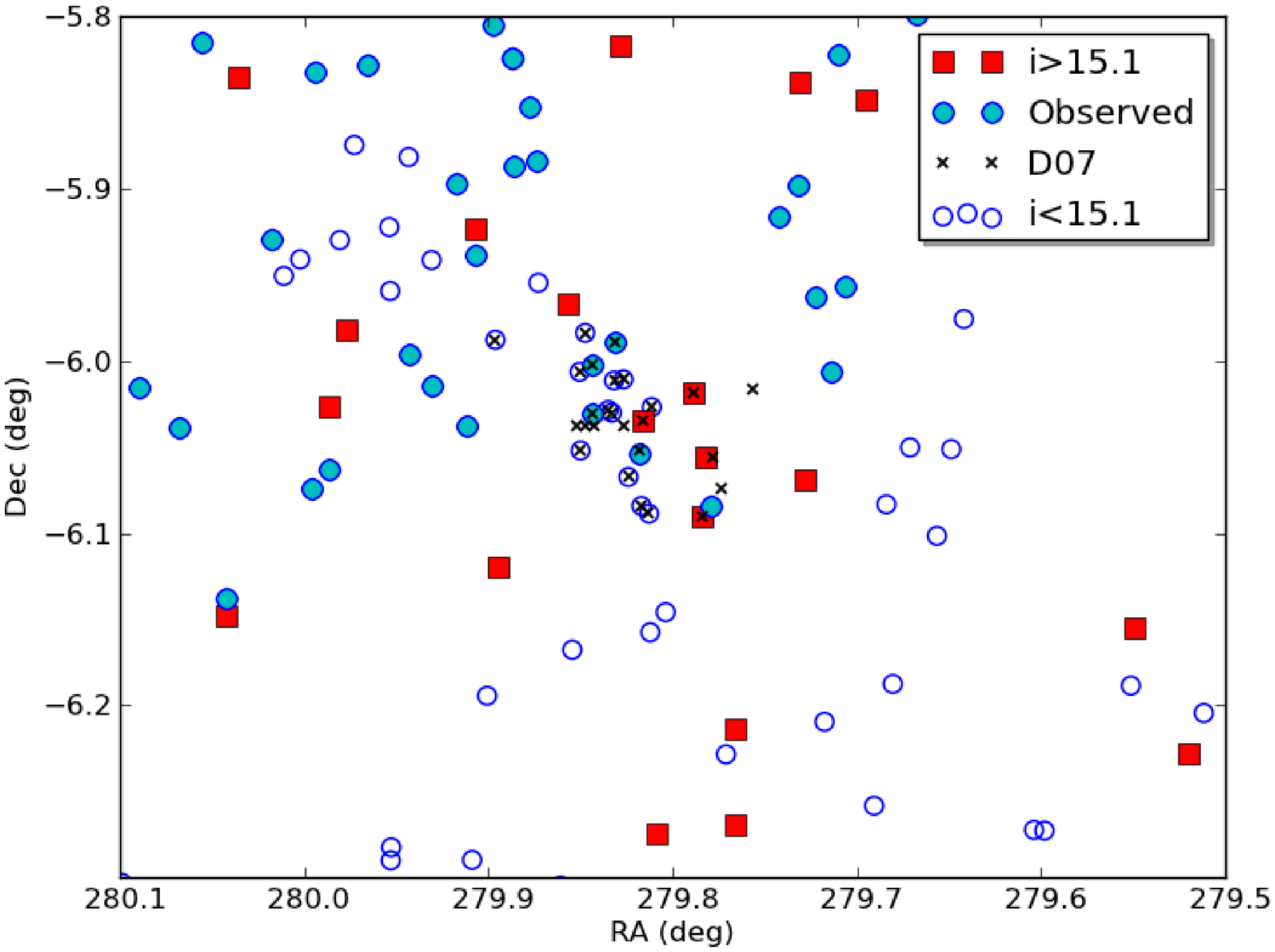}}
\caption{Spatial distribution of candidate red luminous stars, passing all the photometric criteria listed in Section~\ref{sec:target} in the immediate vicinity of Ste~2. Blue circles represent stars with $i<15.1$, while red squares represent candidates with $i>15.1$ or not detected in $i$ (and so too faint to be observed with AF2). Filled circles mark stars actually observed. Members of Ste~2 from \citetalias{davies07} are marked by crosses (note that a few of them have not been selected by our criteria). \label{fig:distall}}
\end{figure}

Our data show that the distribution of RSGs in the vicinity of Ste~2 is very elongated in the NE-SW direction, forming a narrow structure at least $15\arcmin$ long. Is this very strange shape due to an observational bias? This is unlikely, as the DSS2 images suggest the presence of dark clouds precisely in front of the cluster, with significantly lower stellar density than a few arcmin to the East. To investigate this issue, in Fig.~\ref{fig:distall}, we plot all the stars that pass the colour cuts in 2MASS in the immediate vicinity of Ste~2, identifying those that are sufficiently bright in $I$ to be observable in our survey. The distribution of sources fully rules out the possibility that the known membership of Ste~2 is determined by a selection bias. The cluster is surrounded by an area of low density of candidates in all directions except the North East. In addition, none of the stars observed immediately to the West of the cluster has spectral types or $v_{{\rm rad}}$ similar to cluster members (e.g., D12, D38, J18384958$-$0557222), while most of the candidates observed to the NE are red supergiants with $v_{{\rm rad}}$ compatible with an association.

Further to the Northeast, we find the area with a high concentration of RSGs studied by \citet{lopez}, centred close to ($l$\,$=27\fdg0$, $b$\,$=+0\fdg1$). Though these authors do not report accurate coordinates for their objects, we may have some targets in common. For example, TMGS~J184217.8$-$051306 (which they classify M8\,I) is almost certainly 2MASS~J18421794$-$0513098 (M7\,I--II here). Pending the determination of radial velocities for RSGs in this area, it is sensible to assume that at least a fraction of them will have similar colours and reddenings to Ste~2 members.

The \ion{H}{ii} region closest in the sky to Ste~2 is G026.103$-$00.069b, just to the South of the cluster. This region has $v_{{\rm LSR}}$\,$=104\pm2\:{\rm km}\,{\rm s}^{-1}$ \citep{anderson11}. There are several \ion{H}{ii} regions in the neighbourhood with similar velocities, for instance, G025.766+00.211 G025.800+00.240 or G026.555$-$00.305. In principle, these velocities are compatible with that of the cluster. Small offsets between tracers probing different components are expected, as tracers of \ion{H}{ii} generally see the velocity of the expanding gas.
Comparison of velocities derived from H$\alpha$, a tracer of the ionised gas, to velocities derived from CO, a tracer of the cold gas, for a large sample of \ion{H}{ii} regions and associated molecular clouds gives a mean offset $\approx4\:{\rm km}\,{\rm s}^{-1}$ (in the sense of higher CO velocities), with a dispersion of $\approx12\:{\rm km}\,{\rm s}^{-1}$ \citep{fich90}. On the other hand, using radio spectra, \citet{ab09} place all these \ion{H}{ii} regions in the vicinity of Ste~2 beyond corrotation, at a distance $\sim9$~kpc.

Is there a physical relationship between all these objects then? The radial velocity of Ste~2 is compatible with the terminal velocity in this direction. Even though some Galactic radial velocity curves predict higher terminal velocities, tracers that can be seen through very heavy extinction, such as \ion{H}{ii} regions and methanol masers, do not reach higher velocities. In view of this, we cannot guarantee that every single tracer with velocity around $v_{{\rm LSR}}=+110\:{\rm km}\,{\rm s}^{-1}$ is spatially close to Ste~2. Indeed, according to \citet{ab09}, the \ion{H}{ii} regions in its immediate neighbourhood with similar $v_{{\rm LSR}}$ are $\sim$$3$~kpc away.

The RSGs observed in this survey, however, {\em are} affected by extinction. The extinction calculated for several lines of sight (Fig.~\ref{fig:ext}) shows a phenomenal rise at $d\ga6$~kpc, reaching values $E(J-K_{{\rm S}})>2.5$. Our sample only contains one star with comparable reddening, J18424927$-$0445095 with $E(J-K_{{\rm S}})=2.8$. This is a source from the control region with $I=16.1$. The vast majority of sources with radial velocities similar to Ste~2 have $E(J-K_{{\rm S}})<2$, compared to a cluster median value $E(J-K_{{\rm S}})=1.76$.
Therefore none of these RSGs may be behind the cluster by any substantial amount. On the other hand, their LSR velocities force them to be located at distances not much shorter than $\sim6$~kpc. In view of this, we have to conclude that, even though a physical relation (i.e., membership in a single association) cannot be strictly claimed with the data available, all the RSGs that we detect with $v_{{\rm LSR}}\approx+110\:{\rm km}\,{\rm s}^{-1}$ must be concentrated in a small region, unlikely to be more than 500~pc deep.

Further analysis of this issue is complicated by the strong dependence on two basic assumptions:
\begin{enumerate}[i)]
\item Kinematic distances assume no strong deviations from the Galactic rotation curve. Some of our results suggest that this assumption may break down along this line of sight.
\item Other distance estimates to Ste~2 assume an extinction law in the infrared close to the standard law \citep{rieke}. Even though there have been suggestions that this law is almost universal \citep{inde05}, recent work has convincingly shown that the extinction law along high-reddening directions towards the inner Galaxy implies much higher transparency in the $K_{{\rm S}}$ band \citep{nishi08,sh09,gonzalez12}. If this new law was to be adopted, the distance to Ste~2 would certainly increase.
\end{enumerate}

Testing these two assumptions is clearly outside the scope of this paper. Meanwhile, considering the data currently available, we can conclude that Ste~2 consists of a compact cluster with radius between $2\arcmin$ and $2\farcm6$ -- 3.2 to 4.2~pc at 5.5~kpc --, which would contain between 16 and 20 RSGs (depending on the radius chosen), and a much larger halo of stars with similar parameters, a large fraction of which are very likely part of a physical association. Even though we cannot be conclusive about the parameters of the extended association, it likely contains several dozen RSGs.  According to current estimates based on population synthesis models \citep{simonw51}, the central cluster would contain between $\sim3$ and $6\times10^{4}\:M_{\sun}$, and so be gravitationally bound. The association surrounding it would be much more massive, and likely contain many small clusters, which are difficult to identify because of the large number of unrelated RSGs projected along the line of sight. 

Could the whole overdensity of RSGs be a single association? To answer this question, we need first to determine its actual boundaries. With the current data, this possibility cannot be ruled out. The most distant RSGs in our sample, lying $\sim$$90\arcmin$ away from Ste~2, would be located 145~pc away at a distance of 5.5~kpc. Such size is comparable to that of the largest star forming complexes in the Galaxy, such as W43 \citep{nguyen11} and W51 \citep{kumar}. Such complexes, however, do not result in a single starburst of star formation, but rather proceed to sequential (but not necessarily triggered) star formation at multiple sites throughout the cloud \citep[e.g.,][]{simonw51}. Their end products may be large associations, such as the complexes of open clusters in M51 \citep{bastian} or the 30~Dor extended association, surrounding the central starburst cluster R136  \citep{walborn02}, which display a wide age range ($\sim 5$~Myr). 
Conversely, a moderately massive association, such as Per~OB1, with a mass $\approx2\times10^{4}\:M_{\sun}$ in the core region, does not seem to show strong evidence for a significant age spread \citep{currie10}.

With an age $\la20$~Myr, the Ste~2 association has had the time to evolve dynamically. Therefore the actual distribution of RSGs could reflect both a widely multi-seeded star formation process, with many progenitors of RSGs forming in small clusters, or the consequences of dynamical evolution, with many progenitors of RSGs being ejected from Ste~2 and other clusters. If {\it Gaia} could provide us with accurate proper motions for a sizable sample of our RSGs, it could be possible to reconstruct the dynamical evolution of the complex and favour one of the two options.

\section{Conclusions}

In this paper, we have presented the results of a spectroscopic survey of red luminous stars along the Galactic Plane in the $l=26\degr$--$28\degr$ region, covering the vicinity of the starburst cluster Stephenson~2 and a control region $1\fdg1$ away. We have observed $\sim$250~stars amongst $>600$ candidates ($>400$ actually observable with our instrumentation), finding that all except one are red luminous stars. We find $\sim$80 red supergiants, most of M spectral type. This number represents a significant fraction of the RSGs known in the Galaxy, in spite of the fact that we have only surveyed $\sim$1.7~deg$^{2}$, and checked $\sim$40\% of the candidates. The capability to select red luminous stars with very simple colour cuts has been fully demonstrated.

The RSGs found are not strongly concentrated towards Ste~2 (we stress that our observation technique does not allow us to sample but a very small fraction of densely concentrated populations), but seem to be rather uniformly spread over the area surveyed. Moreover, about half the RSGs have radial velocities and reddenings that make a physical connection with the cluster very unlikely.

On the one hand, these findings demonstrate beyond any reasonable doubt that the clusters of RSGs are not isolated, but rather part of extended stellar associations, a result that had already been found for RSGC3 (\citetalias{neg11}; \citealt{carlos}). This diffuse population likely extends over several squared degrees and comprises several hundred objects. Based on their reddenings and radial velocities, members of this population cannot be located at distances from the Earth significantly different from those of the RSGCs. Perhaps they represent the result of starburst caused by the tip of the Galactic bar, as suggested by \citet{lopez}. Wider-field searches can confirm the existence of such a large population and probe its spatial extent.

On the other hand, the high number of RSGs not associated with Ste~2 bears witness to the richness of the young stellar population towards the inner Galaxy. Even though the tip of the Galactic bar may create special conditions at the distance of the Scutum Complex, the foreground population along this sightline is unlikely to be affected. If so, such high numbers could be representative of {\em all} sightlines towards the inner Milky Way. 

Our results show that the limits of Stephenson~2 cannot be clearly defined. A compact core, with radius $\la2\farcm5$ and containing about 20 RSGs, is surrounded by an extended association that merges into a general background with an important over-density of RSGs. To investigate the spatial extent of this overdensity and test whether it covers the whole region containing the clusters of red supergiants ($l\sim 24\degr - 29\degr$), a wider-field spectroscopic survey is needed. Such a survey is currently underway and will be presented in a future paper.

Finally, the technique used here offers a novel way of tracing Galactic structure. Red supergiants are very bright objects in the infrared and can be seen through heavy extinction. Even though some of them may have peculiar velocities, their large numbers and their membership in clusters make them very good tracers. Small clusters containing $\sim$5 RSGs must be relatively common in areas of high young population, and they can serve as very good tracers of Galactic structure. Even if they are too reddened to be accessible around the \ion{Ca}{ii} triplet, they are very good targets for moderately-high-resolution observations in the $K$ band. Investigation of further photometric criteria, making use of a combination of 2MASS and GLIMPSE colours, seems the next natural step along this approach to penetrating into the inner Milky Way.

\begin{acknowledgements}
 
We thank the referee, Dr. Ben Davies, for constructive criticism, leading to substantial improvement of the paper. The WHT is operated on the island of La Palma by the Isaac Newton Group in the Spanish Observatorio del Roque
de Los Muchachos of the Instituto de Astrof\'{\i}sica de Canarias.
The ISIS observations were obtained under the Spanish Instituto de Astrof\'{\i}sica de Canarias Director's Discretionary Time.

This research is partially supported by the Spanish Ministerio de
Ciencia e Innovaci\'on (MICINN) under grants AYA2010-21697-C05-05, AYA2011-24052 and CSD2006-70, and by the Generalitat Valenciana (ACOMP/2012/134). 

 This publication
makes use of data products from 
the Two Micron All 
Sky Survey, which is a joint project of the University of
Massachusetts and the Infrared Processing and Analysis
Center/California Institute of Technology, funded by the National
Aeronautics and Space Administration and the National Science
Foundation.  UKIDSS uses the
UKIRT Wide Field Camera (WFCAM; Casali et al. 2007) and a photometric
system described in \citet{hewett06}. The pipeline processing and
science archive are described in \citet{hambly08}. The DENIS project has been partly
funded by the SCIENCE and the HCM plans of the European Commission
under grants CT920791 and CT940627. 
This research has made use of the POLLUX database
({\tt http://pollux.graal.univ-montp2.fr}) operated at LUPM (Universit\'e
Montpellier II - CNRS, France) with the support of the PNPS and INSU.
This research has made use of the Simbad,
Vizier and Aladin services developed at the Centre de Donn\'ees
Astronomiques de Strasbourg, France.

\end{acknowledgements}

{}

\end{document}